\documentclass[12pt]{iopart}

\expandafter\let\csname equation*\endcsname=\relax 
\expandafter\let\csname endequation*\endcsname=\relax 
\usepackage[utf8]{inputenc}
\usepackage{empheq}
\usepackage{float}
\usepackage{lipsum}
\usepackage{caption}
\usepackage{mathtools,cuted}
\usepackage{esvect}
\usepackage{multirow}
\usepackage{amsfonts, bbm}
\usepackage{xcolor}
\usepackage{soul}
\usepackage[export]{adjustbox}
\usepackage{graphicx, caption}
\usepackage{dcolumn}
\usepackage[none]{hyphenat} 
\usepackage{bm}
\usepackage{hyperref}
\usepackage{physics}
\usepackage[mathlines]{lineno}
\usepackage{cleveref}
\newcommand{\ham}[0]{H}

\usepackage{ulem}

\begin{document}

\title[]{Spectral Properties of Two Coupled Fibonacci Chains}

\author{Anouar Moustaj*\textsuperscript{1}, Malte Röntgen*\textsuperscript{2,3}, Christian V. Morfonios\textsuperscript{2}, Peter Schmelcher\textsuperscript{2,4}, Cristiane Morais Smith\textsuperscript{1}}
\address{\textsuperscript{1}Institute of Theoretical Physics, Utrecht University, Princetonplein 5, 3584CC Utrecht, The Netherlands}%
\address{\textsuperscript{2}Zentrum für Optische Quantentechnologien, Fachbereich Physik, Universität Hamburg, Luruper Chaussee 149, 22761 Hamburg, Germany}
\address{\textsuperscript{3}Laboratoire d’Acoustique de l’Université du Mans, Unite Mixte de Recherche 6613, Centre National de la Recherche Scientifique, Avenue O. Messiaen, F-72085 Le Mans Cedex 9, France}
\address{\textsuperscript{4}The Hamburg Centre for Ultrafast Imaging, Universität Hamburg, Luruper Chaussee 149, 22761 Hamburg, Germany}

\vspace{10pt}
\begin{indented}
    \item \today
\end{indented}    

\begin{abstract}
    The Fibonacci chain, i.e., a tight-binding model where couplings and/or on-site potentials can take only two different values distributed according to the Fibonacci word, is a classical example of a one-dimensional quasicrystal. With its many intriguing properties, such as a fractal eigenvalue spectrum, the Fibonacci chain offers a rich platform to investigate many of the effects that occur in three-dimensional quasicrystals. In this work, we study the eigenvalues and eigenstates of two identical Fibonacci chains coupled to each other in different ways. We find that this setup allows for a rich variety of effects. Depending on the coupling scheme used, the resulting system (i) possesses an eigenvalue spectrum featuring a richer hierarchical structure compared to the spectrum of a single Fibonacci chain, (ii) shows a coexistence of Bloch and critical eigenstates, or (iii) possesses a large number of degenerate eigenstates, each of which is perfectly localized on only four sites of the system. If additionally, the system is infinitely extended, the macroscopic number of perfectly localized eigenstates induces a perfectly flat quasi band. Especially the second case is interesting from an application perspective, since eigenstates that are of Bloch or of critical character feature largely different transport properties. At the same time, the proposed setup allows for an experimental realization, e.g., with evanescently coupled waveguides, electric circuits, or by patterning an anti-lattice with adatoms on a metallic substrate.
\end{abstract}

\vspace{2pc}

%
%
%

\section{\label{sec:Intro}Introduction 
	}
	
  	Aperiodic structures, in particular quasicrystals \cite{shechtman}, have attracted the attention of researchers for many decades \cite{Janssen:a25379, alpdreelec, Vieira2005Low-EnergyChains, Tanese2014FractalPotential, Jagannathan2021RMP93045001FibonacciQuasicrystalCaseStudy}.
	Even the simplest quasiperiodic, one-dimensional models exhibit a rich variety of behaviour, ranging from  critical eigenstate localization properties, to the appearance of energy spectra that sometimes form a fractal set \cite{emaciaaporder}. 
	The tools used to understand these behaviors include, but are not limited to, renormalization procedures, multifractal analysis, or symmetry considerations \cite{tedjanssen, Niu1986PRL572057RenormalizationGroupStudyOneDimensionalQuasiperiodic, Mac__2016}. One of the peculiar features that these systems show is a hierarchical structure of energy gaps \cite{Niu1986PRL572057RenormalizationGroupStudyOneDimensionalQuasiperiodic}. This gives rise to spectral measures that are singular continuous, such as the fractal set observed in the Fibonacci quasicrystal \cite{Jagannathan2021RMP93045001FibonacciQuasicrystalCaseStudy}. This is a feature that was linked to all the critical localization properties observed in quasicrystals as well as other systems that possess some type of aperiodic order \cite{emaciaaporder}.
	
	Although the behavior of aperiodic chains has been investigated extensively and in great detail, comparatively little work has been dedicated to the case where two or more chains are coupled to each other, forming an \textit{aperiodic ladder} \cite{Moreira2006SpecificSequences, Pal2014AbsolutelyNetworks, Mukherjee2017EPJB9052ControlledDelocalizationElectronicStates,Saha2019ParticleLadder,Roy2022LocalizationEdge}. In this work, we take a step into this realm by analyzing a range of different coupling schemes between two identical one-dimensional Fibonacci chains. Specifically, we study cases where the two chains are directly coupled in a uniform, non-uniform, or quasiperiodic manner, and evaluate the resulting spectral properties, namely the energy eigenvalues and eigenstates. Additionally, we study a special case of an indirect coupling, that is, two chains coupled to each other through some intermediate sites. These cases are easily tractable, since they posses a reflection symmetry which allows to block-diagonalize the Hamiltonian.
	
	We find different spectra depending on the setup. In the case of uniform coupling, the eigenvalue spectrum is identical to that of two uncoupled Fibonacci chains, but with shifted energy eigenvalues. On the other hand, if the two chains are coupled only through a single site, the spectrum consists of two Fibonacci chains with an on-site defect. The structure of the eigenvalue spectrum becomes more complex for the case of quasiperiodic coupling, for which a richer hierarchical structure reveals itself through a perturbative renormalization approach. If only the sites of one specific type (A or B) are coupled to each other, we show that, for a specific value of the interchain coupling, half of the eigenstates are critical, while the other half are extended. Interestingly, these two different classes of eigenstates possess different parity with respect to a corresponding reflection operation and can thus be selectively excited by incoming waves of negative or positive parity.
    This could be used to control the transport properties of this system.
    Finally, we also realise couplings between the chains through some intermediate sites. This leads to the appearance of flat bands in a quasiperiodic lattice.
    
	This paper is organized as follows. To be self-contained, we start by briefly reviewing the properties of the Fibonacci chain in \Cref{sec:fiboRing}, followed by an overview of the methods used to generate our results in \Cref{sec: Methods}.
	In \Cref{sec:uniformCoupling}, we analyze the simplest way of coupling two Fibonacci chains, namely a uniform one.
	In \Cref{sec:non-UniformCoupling}, we investigate different cases of non-uniform coupling.
	We start by connecting only sites of a specific type to each other in \Cref{subsec:AOrBCoupling}.
	Then, in \Cref{subsec:defCoupling}, we analyze the case where only two sites are coupled.
	In \Cref{sec:quasiPeriodicCoupling}, we couple the two chains in a quasiperiodic fashion and analyze the resulting eigenvalue spectrum in terms of a renormalization scheme.
	Finally, in \Cref{sec:IntermediateCoupling}, we consider the scenario where the two chains are not directly coupled to each other, but through intermediate sites.
	Our conclusions are presented in \Cref{sec:Conclusion}.


\section{A single Fibonacci chain} \label{sec:fiboRing}
	We consider a general Fibonacci chain model, namely a nearest-neighbour tight-binding chain with periodic boundary conditions, so that the chain effectively becomes a ring.
	The on-site potential and hopping amplitudes are both modulated by the Fibonacci sequence, and the corresponding Hamiltonian is given by 
	\begin{equation}\label{genfibo}
	\ham_{F} = \sum_{i=1}^{N} v_{i} \ket{i}\bra{i} + \sum_{\langle i,j \rangle}h_i \ket{i}\bra{j}
	\end{equation}
	where $\ket{i}$ denotes a basis state fully localized on the $i$-th site and $\langle i,j \rangle$ denotes nearest-neighbors. The on-site potential $v_i$ and the hopping amplitude $h_i$ are both binary and follow the sequence of a Fibonacci word $S_n$ of generation $n$. 
	The latter is defined by the recursion relation 
	\begin{equation*}
	    S_n=S_{n-1}S_{n-2}, \ \ n\geq 2,
	\end{equation*}
	which is a binary representation of the Fibonacci sequence. This inductive recursion formula is expressed as a string concatenation instead of number addition. Another typical representation is through the substitution rule $A\to AB$ and $B\to A$. With the initial two words being $S_{0}=B$, $S_{1} = A$, the first few words are thus
	\begin{align*}
	S_{0} &= B \\
	S_{1} &= A \\
	S_{2} &= AB \\
	S_{3} &= ABA \\
	S_{4} &= ABAAB \\
	S_{5} &= ABAABABA \\
	S_{6} &= ABAABABAABAAB \, .
	\end{align*}
	We note that the Fibonacci sequence is most commonly represented in numeric form through the recursion formula
	\begin{equation*}
	    F_n= F_{n-1}+F_{n-2}, \ \ n\geq 2,
	\end{equation*}
	where $F_{0} = F_{1} = 1$ and $F_n$ is the $n^\text{th}$ Fibonacci number, which is equal to the word length $|S_n|=F_n$. Another important property of the sequence is that, in the thermodynamic limit, the ratio between the amount of letters A and B is equal to the golden ratio $\phi$. This is more appropriately expressed as \cite{KILIC2008701}
	\begin{equation*}
	    \lim_{n\to\infty}\frac{F_{n}}{F_{n-1}}=\frac{1+\sqrt{5}}{2}\equiv\phi \, .
	\end{equation*}
	
	The Hamiltonian \eqref{genfibo} can be studied in its general form or reduced to a purely on-site or hopping Fibonacci model, where the parameters $h_i=h$ or $v_i=v$ are uniform. These have been termed diagonal and off-diagonal models in the literature, while \Cref{genfibo} is referred to as the mixed model. All these models have been studied previously in their various forms \cite{SireMosseri, NORI, Kohmoto1986PRB34563QuasiperiodicLatticeElectronicProperties, Kohmoto1987PRB351020CriticalWaveFunctionsCantorset}, but one important aspect is the equivalence between the on-site and hopping models under a perturbative renormalization scheme \cite{NORI}. This means that we can uncover all essential features by studying the unmixed models. For this reason, we will focus on the on-site model for our analysis.
	In this case, the on-site potential $v_{i}$ of the $i$-th site is equal to either $v_{A}$ or to $v_{B}$, depending on whether the $i$-th character of $S_{n}$ is equal to $A$ or to $B$, and the hopping parameter is constant and equal to $h$. We note that the number of sites in the chain is then equal to $|S_n|=F_n$.
 \begin{figure}[!t]
\centering
\includegraphics[max width = 0.5\textwidth]{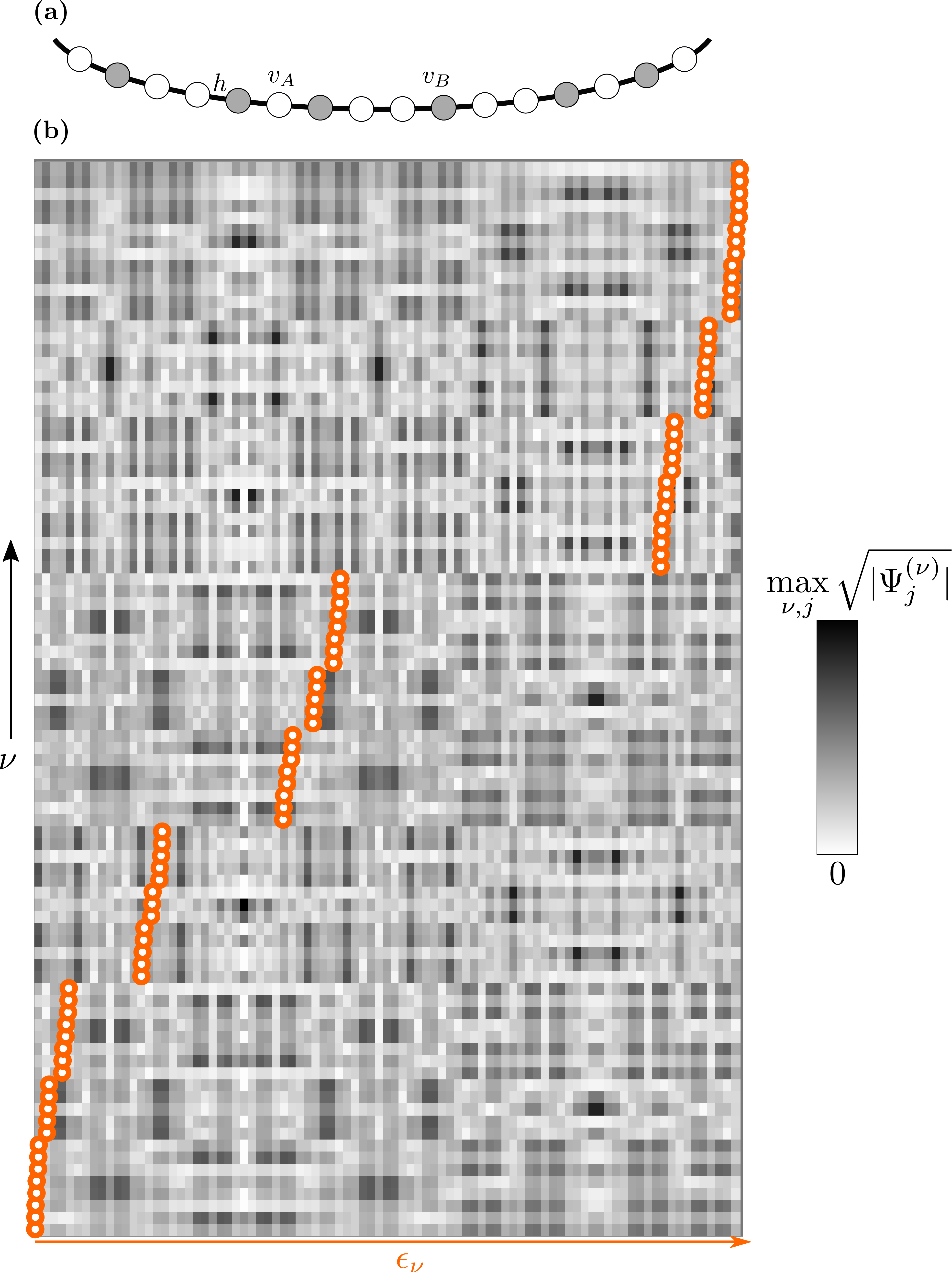}
\caption[]{(a) Graphical representation of a Fibonacci chain with periodic boundary conditions (see text for details). (b) Eigenstate map for such a Fibonacci chain of $N = 89$ sites.
		Each of the $N$ rows of the gray scale image shows the value of $|\Psi^{(\nu)}(j)|^{1/2}$ (with each eigenstate normalized), where $j$ is a site index, $\nu$ is an energy index, and with each square pixel corresponding to one of the $N$ sites.
		The orange dots denote the eigenenergy $\epsilon_{\nu}$ of the eigenstates. The eigenvalues and eigenstates were obtained for $h=-1$, $v_A = 0$ and $v_B = 3$.
	}
\label{fig:isolatedRing}
\end{figure}
	\Cref{fig:isolatedRing} (a) shows a graphical representation of such a Fibonacci chain.
	Due to the periodic boundary conditions, the depicted system constitutes the unit cell of a periodic approximant of the quasicrystal (if $N\to\infty$, it becomes a proper quasicrystal). This is a system which has long-range order, without being periodic. Its eigenstates form a set of critical states, which have atypical localization properties. This means that they are neither extended nor localized \cite{emaciaaporder}. The eigenvalues have a spectral measure that is singular continuous and form a fractal set (a Cantor set of measure zero). This feature is also observed in the eigenstates, with wavefunctions that show multifractal properties \cite{Mac__2016}.
	In \Cref{fig:isolatedRing} (b), the eigenstates of a Fibonacci chain with $N = 89$ are graphically depicted.
 
	The properties of quasicrystals have been studied in a multitude of ways, ranging from perturbative methods based on a renormalization formalism \cite{NORI,Niu1986PRL572057RenormalizationGroupStudyOneDimensionalQuasiperiodic} to exact results using a transfer matrix approach \cite{Kohmoto1983PRL501870LocalizationProblemOneDimension} or a symmetry perspective, which offers insights on the fragmentation of states in terms of local spatial structures of the chain \cite{Rontgen2019PRB99214201LocalSymmetryTheoryResonator}.

    Before we continue, let us briefly comment on possible realizations of the Hamiltonian \cref{genfibo}.
    We note that the following statements also hold for the more complex models that we will present later in this manuscript.
    Since we treat the matrix eigenvalue problem $H \ket{\Psi} = E \ket{\Psi}$, the setups proposed in this work are well suited for a realization in various systems modelled by such a problem, as long as the relevant matrix elements that correspond to couplings and on-site potentials of our Hamiltonian can be controlled.
    An immediate candidate, for instance, are systems of evanescently coupled optical waveguides \cite{Szameit2012DiscreteOpticsFemtosecondLaser}. 
    Here, each site corresponds to a monomodal waveguide.
    By suitably tuning quantities such as waveguide shape, spacing, the on-site potentials $v_i$ and couplings $h_{i,j}$ are tunable in a wide range.
    We remark that the propagation of light in such arrays of evanescently coupled photonic waveguides can be described by a discrete, time-dependent Schrödinger equation, which would also allow to probe the dynamics of the setups proposed in this work.
    Another realization that is directly feasible is in terms of electric circuits, where the sites are nodes of the circuit, and with on-site potentials and couplings being determined by how the nodes are grounded and interconnected to each other \cite{Lee2018CP11TopolectricalCircuits,Dong2021PRR3023056TopolectricCircuitsTheoryConstruction}.

    \section{Methods}\label{sec: Methods}
    In this section, we will provide a comprehensive description of the methods employed to calculate and illustrate the relevant quantities before delving into the obtained results. We will begin by explaining the numerical methods utilized in this study, which encompass the creation of all eigenstate maps. Subsequently, we will outline the analytic methods employed to describe the section concerning quasiperiodic coupling. 

    \subsection{Numerical methods}
    Unless mentioned otherwise, all eigenvalues and eigenvectors were obtained by numerically diagonalizing the corresponding Hamiltonian. In the case of two coupled Fibonacci chains, we individually diagonalized the two Hamiltonians $\ham_{\pm}$ (see \cref{sec:uniformCoupling}) and then constructed the total eigenstates by symmetrizing/anti-symmetrizing these states. We note that this procedure automatically provides an assignment of the eigenvalues of the total Hamiltonian to negative/positive parity.
    The eigenstate maps (as in \cref{fig:isolatedRing}) and energy plots (as in \cref{fig:uniformCoupling_1}) were produced with Mathematica.

    \subsection{Analytical methods}
    \paragraph{Hierarchical splitting and renormalization}
    The main analytic tool used in this work is a perturbative renormalization procedure based on the degenerate Brillouin-Wigner perturbation theory, described in the works of Niu and Nori \cite{Niu1986PRL572057RenormalizationGroupStudyOneDimensionalQuasiperiodic, NORI}. 
    \begin{figure}[!hbt]
        \centering
        \includegraphics[width=0.9\textwidth]{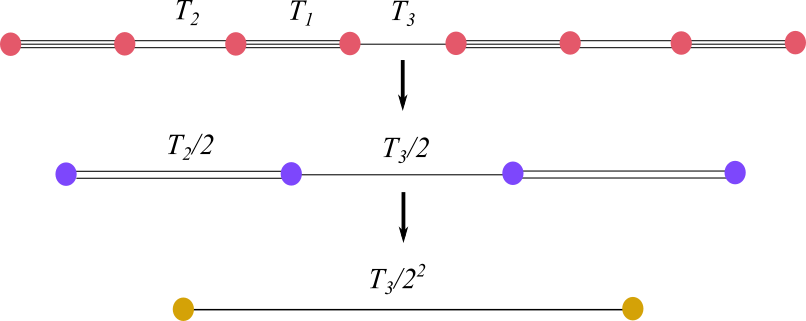}
        \caption{Graphical representation of the renormalization procedure. The topmost chain has a hierarchical distribution of hoppings, with $T_3\ll T_2 \ll T_1$. The blue sites in the middle chain represent the (anti)bonding states of the unperturbed Hamiltonian of the topmost chain, which have effective couplings $T_2/2$ and $T_3$/2. In its turn, the lowermost chain represents the (anti)bonding states of the middle chain, with its effective coupling given by $T_3/2^2$. Figure inspired from \cite{Niu1986PRL572057RenormalizationGroupStudyOneDimensionalQuasiperiodic}.}
        \label{fig: Hierarchical hopping and renormalization}
    \end{figure}
    
    Consider a one-dimensional chain described by a tight-binding Hamiltonian that incorporates a set of hierarchical hoppings, where the hopping strengths follow the relation $T_j\ll T_{j-1}\ll \cdots T_{2}\ll T_1$ (as depicted in the topmost chain of \Cref{fig: Hierarchical hopping and renormalization}). In this scenario, we can employ a perturbative renormalization approach to calculate the energy levels of the topmost chain.

    This perturbative renormalization approach involves treating the weaker hoppings ($T_j$ with higher indices) as perturbations to the dominant hopping term ($T_1$). This allows us to systematically incorporate the effects of the weaker hoppings and compute the resulting energy levels of the topmost chain. This is done by using the subset of unperturbed degenerate eigenstates as the basis set for the perturbed Hamiltonian. We then apply Brillouin-Wigner perturbation theory to calculate the effective couplings between these states. In \Cref{fig: Hierarchical hopping and renormalization}, this process is visually represented by the middle chain, where each blue site represents a (anti)bonding degenerate eigenstate of the $T_1$ molecule. By considering these (anti)bonding states as the basis, we can determine the effective couplings induced by the weaker hoppings ($T_j$) and calculate the resulting energy levels of the topmost chain. In principle, this process is repeated indefinitely, but in practice, one takes a chain of finite size and imposes periodic boundary conditions to perform the calculations. The energy levels form a cluster of $2^j$ values around the base energy $E_0$
    \begin{equation}
        E=E_0\pm T_1\pm\frac{T_2}{2}\pm\cdots\pm\frac{T_j}{2^{j-1}} \,.
    \end{equation}
    
    \paragraph{Brillouin Wigner perturbation theory} We shall now give a brief overview of Brillouin-Wigner perturbation theory. Consider a Hamiltonian $H=H_0+H_1$, with $H_1$ acting as a perturbation to $H_0$. Let $E_0$ be a degenerate energy eigenvalue, and $Q$ denote the projection operator onto the corresponding degenerate subspace. We also denote the complementary projection operator $P=\mathbbm{1}-Q$. The eigenvalue equation $H\ket{\psi}=E\ket{\psi}$ can be rewritten such that 
    \begin{equation}\label{Eq: psi projected out}
        P\ket{\psi}=P\frac{1}{E-H_0}H_1\ket{\psi}=P\frac{1}{E-H_0}PH_1\ket{\psi}
    \end{equation}
    where the last equality follows from $P^2=P$ and $PH_0=H_0P$. We now write
    \begin{align*}
        \ket{\psi}&=(Q+P)\ket{\psi} \\
        &=Q\ket{\psi}+P\frac{1}{E-H_0}PH_1\ket{\psi} \\
        &=Q\ket{\psi}+P\frac{1}{E-H_0}PH_1(Q+P)\ket{\psi} \\
        &\vdotswithin{=} \\
        &=\sum_{n=0}^\infty\left(P\frac{1}{E-H_0}H_1\right)^nQ\ket{\psi},
    \end{align*}
where we have just consistently used \Cref{Eq: psi projected out}. With the above equation, we can now easily obtain an effective Hamiltonian for the degenerate subspace with eigenvalue $E_0$ through a left multiplication by $QH$, 
\begin{equation}\label{Eq: Effective Ham Deriv}
\begin{split}
    QH\ket{\psi}=EQ\ket{\psi}&=\left[QH_0Q+QH_1\sum_{n=0}^\infty\left(P\frac{1}{E-H_0}H_1\right)^nQ\right]Q\ket{\psi} \\
    &\equiv H_{\text{eff}}Q\ket{\psi}
\end{split}
\end{equation}

Using this effective Hamiltonian, it will be possible to understand the structure of the spectrum in the case of quasiperiodic coupling, which is treated in \Cref{sec:quasiPeriodicCoupling}.

\section{Uniform coupling} \label{sec:uniformCoupling}
\begin{figure}[!hbt]
		\centering
		\includegraphics[max width =\textwidth]{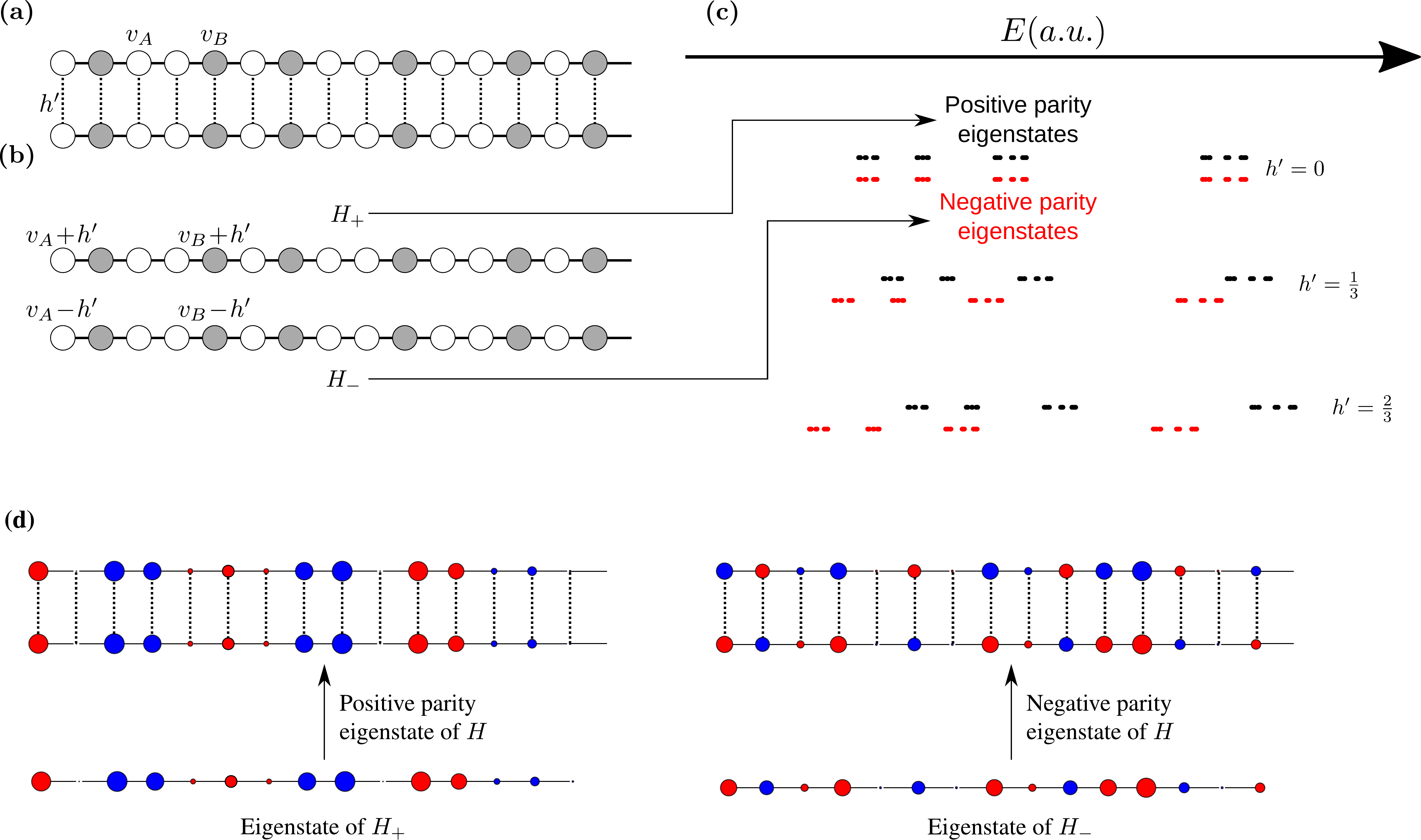}
		\caption{(a) Two uniformly coupled Fibonacci chains.
		(b) The result of writing the original system in terms of the symmetry-adapted basis \Cref{eq:symBasPos,eq:symBasNeg}. In this basis, the setup decomposes into the two disconnected chains $H_{+}$ and $H_{-}$. (c) A sketch of the shifting spectra for different coupling strengths $h'$ (see text for details). The eigenstates of $H_+$ and $H_-$ correspond to eigenstates of the total Hamiltonian $H$ with positive and negative parity, respectively represented in (d). Here, specific eigenstates of $H_+$ and $H_-$ are shown. At each site, the sign of the eigenstate is represented by red/blue color, while the amplitude is depicted by the radius of the circle.}
		\label{fig:uniformCoupling_1}
	\end{figure}
	Let us investigate the setups where two such chains---each consisting of $N$ sites and with periodic boundary conditions---are coupled to each other in different ways.
	We will focus on the impact of these different coupling schemes on the corresponding spectral properties.

	The first and simplest scenario occurs when the two chains are uniformly coupled to each other, as shown in \Cref{fig:uniformCoupling_1} (a).
	The setup is then described by
	\begin{equation} \label{uniformCoupling}
	\ham = \ham_{I} + h' \sum_{i=1}^{N} \Big(\ket{i_{u}}\bra{i_{l}} + \ket{i_{l}}\bra{i_{u}} \Big)
	\end{equation}
	with
	\begin{equation} \label{eq:isolatedHam}
	\ham_{I} = \sum_{x=u,l}\sum_{i=1}^{N} v_{i} \ket{i_{x}}\bra{i_{x}} + h \sum_{x=u,l}\sum_{\langle i,j \rangle} \ket{i_{x}}\bra{j_{x}}
	\end{equation}
	where $\ket{i_{u}}$, $\ket{i_{l}}$ denote basis states fully localized on the $i$-th site of the upper or lower chain, respectively.
	We note that such a setup has been investigated in Ref.\cite{Lazo2000IaNSV387MultifractalBehaviorFibonacciCrystal}, though with a focus on the density of states and not on the eigenvalues and eigenstates.

	To understand the impact of such a uniform coupling, we employ the up/down mirror symmetry of the setup.
	Due to this symmetry, the eigenstates have definite parity under an exchange of the lower and upper chain. This fact can be used to construct a symmetry-adapted basis $\mathcal{S}$, consisting of $N$ states of positive parity
	\begin{equation} \label{eq:symBasPos}
	\ket{1_{u}} + \ket{1_{l}}, \ldots{}, \ket{N_{u}} + \ket{N_{l}}
	\end{equation}
	and $N$ states with negative parity
	\begin{equation} \label{eq:symBasNeg}
	\ket{1_{u}} - \ket{1_{l}}, \ldots{}, \ket{N_{u}} - \ket{N_{l}} \, .
	\end{equation}
	Written in this basis, $\ham' = \mathcal{S}^{-1} \ham \mathcal{S}$ consists of two isolated subsystems, $\ham_{+}$ and $\ham_{-}$.
	These two subsystems are shown in \Cref{fig:uniformCoupling_1} (b).
	It can be seen that each of them is equal to an isolated Fibonacci chain of $N$ sites, though with on-site potentials uniformly shifted by an amount of plus or minus $h'$.
	That is, $\ham_{\pm} = \ham_{I} \pm h' I$, with $I$ being the identity matrix.
	
	Let us now explore the implications of the above, starting with the eigenvalues of $\ham$.
	Since $\ham$ and $\ham'$ are related by a similarity transformation, the two Hamiltonians share the same eigenvalue spectrum.
	Moreover, since $\ham'$ consists of the two disconnected chains $\ham_{\pm}$, the eigenvalue spectrum of $\ham'$, $\sigma(\ham')=\sigma(H)$, is given by the combination of the eigenvalue spectra of these chains \footnote{To be precise, the spectrum of $\ham$ is the multiset sum of the spectra of $\ham_{+}$ and $\ham_{-}$.}.
	Now, because $\ham_{\pm} = \ham_{I} \pm h' I$, we see that the eigenvalue spectrum of $\ham_{+}$ ($\ham_{-}$) is that of $\ham_{I}$ shifted upwards (downwards) by $h'$, respectively.
	In other words, the inter-chain coupling strength $h'$ plays the role of an ``energy shift parameter'' [see \Cref{fig:uniformCoupling_1} (c)].
	Before we continue, we remark that the eigenstates of $\ham$ can be simply constructed from those of $\ham_+$ and $\ham_-$ by symmetrizing or anti-symmetrizing these states; this is demonstrated in \Cref{fig:uniformCoupling_1} (d).
	
	\section{Non-uniform coupling} \label{sec:non-UniformCoupling}

     \begin{figure*}[!hbt]
    \centering
    \includegraphics[max width=\textwidth]{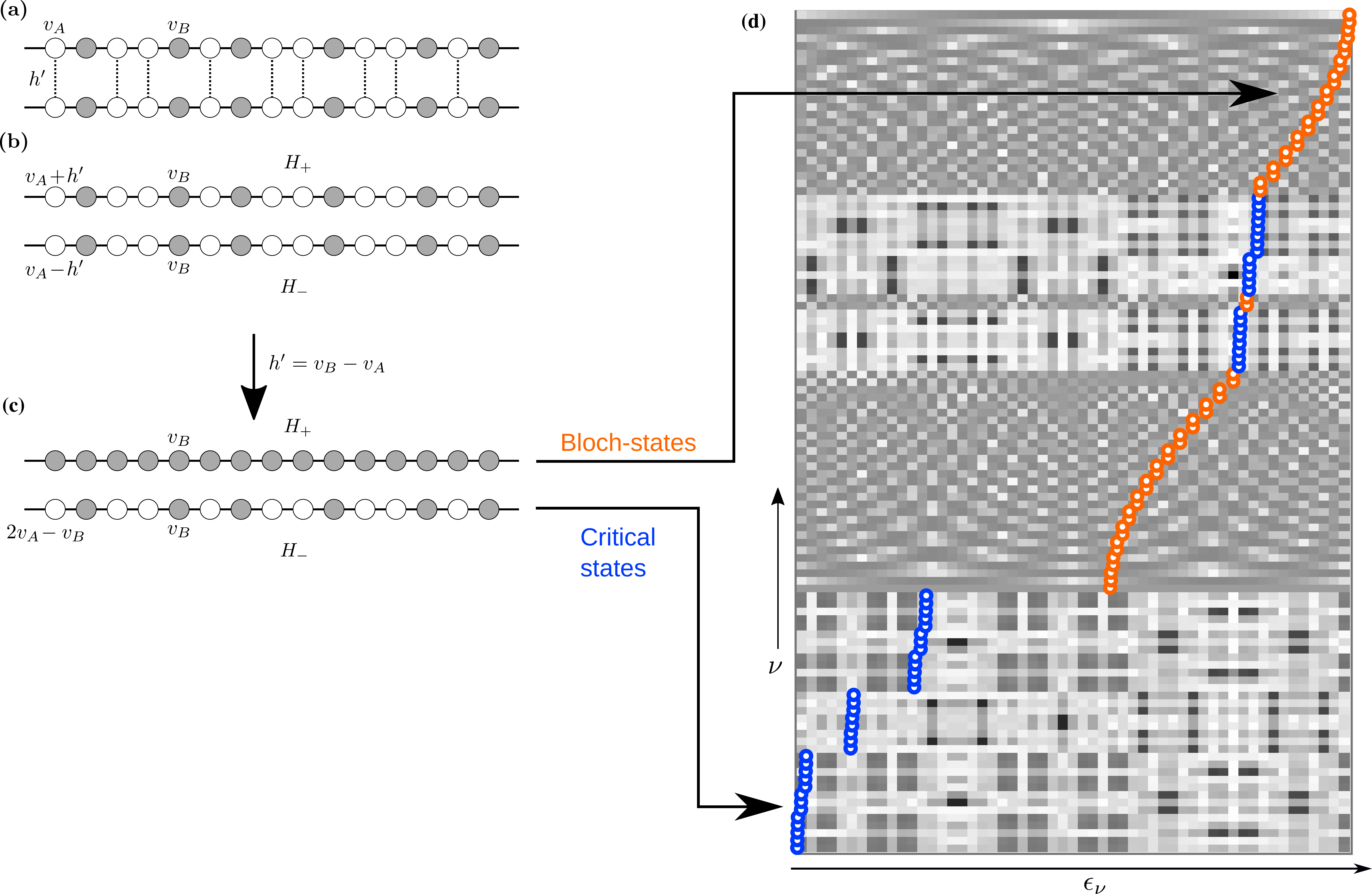}
    \caption{
    (a) Two Fibonacci chains that are coupled by connecting their $A$-sites to each other through couplings with strength $h'$.
    (b) The decomposition into $H_{+}$ and $H_{-}$. (c)
    When choosing $h' = v_B - v_A$, $H_+$ becomes periodic, while $H_-$ remains a Fibonacci chain. Thus, the positive parity eigenstates of $H$ (which correspond to symmetrized eigenstates of $H_+$) are of Bloch-character, while the negative parity eigenstates of $H$ (which correspond to anti-symmetrized eigenstates of $H_-$) are critical states. (d) The eigenstate map of the setup for this particular choice of $h'$. Since the eigenstates $\Psi^{(\nu)}$ of $H$ are either symmetric (orange dots; Bloch-states) or anti-symmetric (blue dots; critical states), each of the $2N$ rows in this eigenstate map only shows the amplitudes on one half of the sites of the total system; the total eigenstate could be obtained through symmetrization or anti-symmetrization.
    The eigenvalues and eigenstates were obtained for $h=-1$, $v_a = 0$ and $v_B = 3$. The color map is the same as in \cref{fig:isolatedRing}, with black pixels corresponding to $\max\limits_{\nu,j} \sqrt{|\Psi_j^{(\nu)}|}.$
    }
    \label{fig:onlyACoupling2}
    \end{figure*}
	Having understood the impact of coupling the two chains uniformly, we now proceed to more complex scenarios. In all cases, we will maintain the reflection symmetry between the two chains.
	Thus, we can still decompose the total Hamiltonian into two smaller chains $H_+$ and $H_-$.
	
	\subsection{Coupling only $A$ or only $B$-sites} \label{subsec:AOrBCoupling}
		
	In the first case, we couple only the $A$ sites to each other, as depicted in \Cref{fig:onlyACoupling2}(a).
	Repeating the same steps as above, we obtain $\ham' = \mathcal{S}^{-1}\ham \mathcal{S} = \ham_{+} \oplus \ham_{-}$, though now with 
	\begin{equation} \label{couplingOnlyAOrB}
	\ham_{\pm} = \sum_{i=1}^{N} v^{\pm}_{i} \ket{i}\bra{i} + h \sum_{\langle i,j \rangle} \ket{i}\bra{j}
	\end{equation}
	where $v^{\pm}_A=v_A\pm h'$, while $v_B^{\pm} = v_B$ is unchanged [see \Cref{fig:onlyACoupling2}(b)].
	In a completely analogous manner, coupling only the $B$-sites to each other will result in an energy shift of the on-site potentials of the $B$-sites only.
	
	A particularly interesting case occurs when coupling only the $A$-sites and setting $h' = v_B - v_A$.
    For this special choice of $h'$, $\ham_{+}$ becomes a uniform chain with zero on-site potential.
	However, $\ham_{-}$ is still a Fibonacci chain.
	Now, since the eigenstates of $\ham_{\pm}$ correspond to positive/negative parity eigenstates of the full chain, the system features an interesting combination of traits: while the positive parity eigenstates are \textit{extended}, the negative parity eigenstates are \textit{critical}. 
    For a quantum system, this means that the phase diagram of such a Hamiltonian features a special point $h' = v_B - v_A$ , at which the system's behavior is highly dependent on the energies of one-particle excitations. As depicted in \Cref{fig:onlyACoupling2}(a), depending on the energy of these parity eigenstates, they will either form Bloch waves (orange energy levels), or critical states (blue energy levels). It is interesting to see that a specific point in parameter space shows a mixture of singular continuous and absolutely continuous spectra.
    This provides a platform where both properties of extended and critical states can be exploited by tuning the Fermi level. In the critical regime, for example, (thermal) conductivities are in general very low (in some cases, they are even lower than for conventional insulators) \cite{archambault_janot_1997, refId0}. On the other hand, the fully extended regime provides the possibility to have phases with high (electrical) conductivities.
    
    Finally, let us note that the possibility of Bloch-states in coupled aperiodic setups has also been observed in Refs. \cite{Pal2014AbsolutelyNetworks, Mukherjee2017EPJB9052ControlledDelocalizationElectronicStates}, in which more complicated coupling schemes have been used.

	
	\begin{figure}
		\centering
  		\includegraphics[max width = 0.5\textwidth]{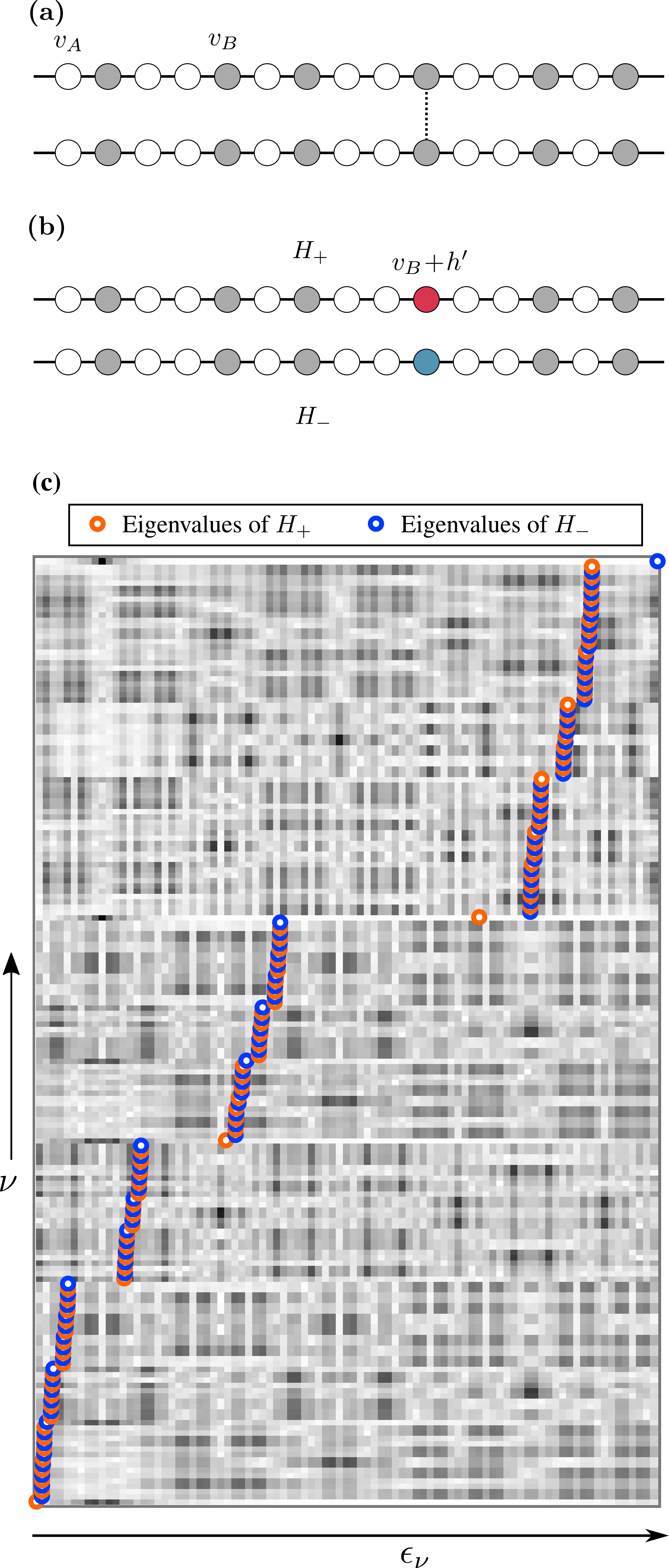}
		\caption{(a) Two Fibonacci chains coupled by connecting a single site of each chain to each other.
		(b) The decomposition into $H_{+}$ and $H_{-}$.
		(c) Eigenstate map for a setup of two Fibonacci chains that are coupled together at the $10$-th site.
		As a result of this \emph{defect-coupling} (see text for details), several eigenstates lie in gaps between quasi bands.
		Orange dots correspond to eigenvalues of $H_{+}$ and blue dots to those of $H_{-}$.
	    The eigenvalues and eigenstates were obtained with $h=-1$, $v_A = 0$ and $v_B = 3$. The color map is the same as in \cref{fig:isolatedRing}, with black pixels corresponding to $\max\limits_{\nu,j} \sqrt{|\Psi_j^{(\nu)}|}.$
		}
		\label{fig:Perturbations}
	\end{figure}
	
	\subsection{Defect-coupling} \label{subsec:defCoupling}
		
		\begin{figure}
	    \centering
	    \includegraphics[max width = 0.5\textwidth]{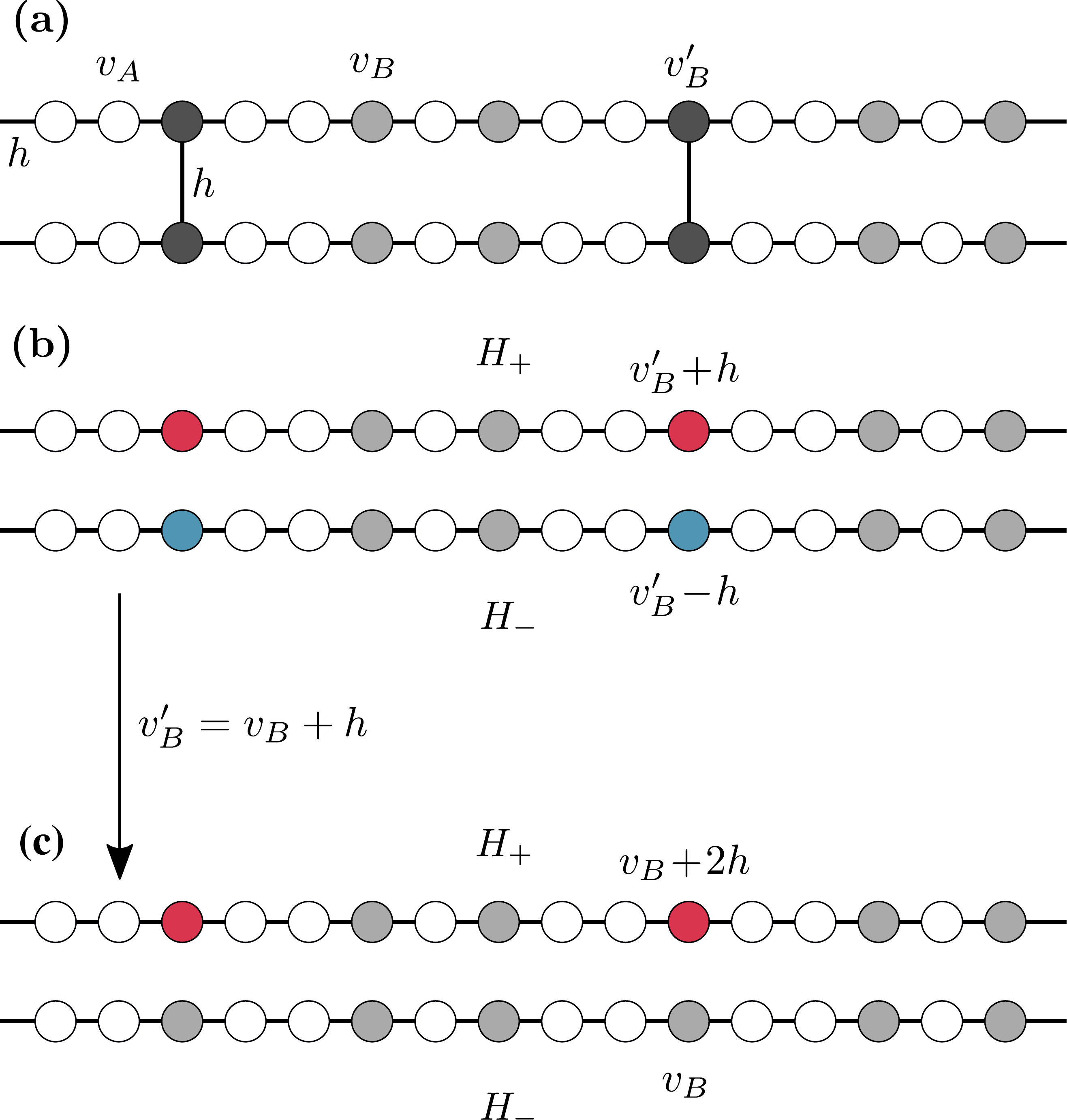}
	    \caption{(a) An excerpt of two Fibonacci chains coupled to each other in a quasiperiodic manner, that is, by coupling the $B$-sites occuring in the pattern $AABAA$ together.
	    (b) The result of the symmetry-induced decomposition into $H_+$ and $H_-$. (c) Since we know that the symmetry-induced decomposition results in a different on-site potential for $H_\pm$, we already assign a different $v_B'$ to the coupled sites, which for ease of analysis is set to $v_B'=v_B+h$, resulting in the case we study in more detail in \Cref{subsec: Decimation Procedure}. In this simplified version, $H_-$ is the standard Fibonacci chain, so we only need to analyze $H_+$. 
	    }
	    \label{fig:quasiperiodicoupling}
	\end{figure}

	
	Another possibility is the selective coupling of only a small subset of sites.
	In the extreme case, this subset consists of one site in each chain [see \Cref{fig:Perturbations} (a)]. The result of such a coupling can be easily deduced. Using the symmetry adapted basis, we see that $H_\pm$ will both be simple Fibonacci chains, with added impurities at the sites that are coupled [see \Cref{fig:Perturbations} (b)]. Such systems have been analyzed previously, and it was found that a single weak impurity is sufficient to render the spectrum unstable and reduce its fractal dimension, leading to a loss of criticality in all states \cite{subsdisorder}. On top of that, it was also recently found that this does not affect all states equally. Using Niu's renormalization procedure \cite{Niu1986PRL572057RenormalizationGroupStudyOneDimensionalQuasiperiodic,Niu1990PRB4210329SpectralSplittingWavefunctionScaling}, it was shown that the degree of criticality loss is dependent on the renormalization path of the site at which the impurity is placed \cite{Moustaj2021PRB104144201EffectsDisorderFibonacciQuasicrystal}.
	This means that the location of the impurity impacts which states of the unperturbed Fibonacci chain are the most affected.
	
	Alternatively, the impact of a single-site defect can be analyzed in a framework of local resonators \cite{Rontgen2019PRB99214201LocalSymmetryTheoryResonator}. These local resonators form building blocks of the whole chain. In an unperturbed (without impurity) chain, each eigenstate is approximately symmetric with respect to a local parity operator. The sites with the highest amplitudes will be the ones corresponding to the resonator structures, which are symmetric under local parity. This approximate symmetry depends on the contrast $c=|h|/|v_A-v_B|$, and is exact in the limit $c\to0$. By placing an impurity on a particular site, one creates a new local resonator structure in the chain, and as such, new localization properties arise, yielding states with amplitude distributions that are radically different from the rest of the eigenstates. This can be seen in \Cref{fig:Perturbations} (c), where the in-gap states have very strong localization, marked by the darker patches around a particular region of the chain (see the topmost level, for example, where there is a very dark patch around site $j=10$) around this new local resonator block \cite{Rontgen2019PRB99214201LocalSymmetryTheoryResonator}.

	\section{Quasiperiodic coupling} \label{sec:quasiPeriodicCoupling}

	Yet another alternative way of coupling the two chains is in a quasiperiodic manner.
	Out of the many possibilities, here we illustrate an immediate and interesting one: We couple only a subset of $B$-sites to each other; namely, those appearing in the pattern $AABAA$, as shown in \Cref{fig:quasiperiodicoupling}(a), where the darker $B$ site sits in between two $A$ sites on each sides. We further set the coupling between the chains to $h$ and choose the coupled $B$ site energy to $v_B'=v_B+h$, such that $H_{-}$ becomes a regular Fibonacci chain, with on-site energies $v_A$ and $v_B$. On the other hand, $H_{+}$ now features new on-site energies: $v_B+2h$, distributed in a quasiperiodic manner. For this choice of coupling, there are $F_{N-5}$ such new sites (which is the number of $AABAA$ blocks in a chain of length $F_N$). 
 
    In analyzing the effective chain using a renormalization approach, we make the assumptions that $h<0$ and $v_A<v_B$. These assumptions are made to enforce the trifurcating structure necessary for the Fibonacci chain's renormalization group flow.
	
	\subsection{Decimation Procedure.}\label{subsec: Decimation Procedure}
	\begin{figure}
		\centering
		\includegraphics[max width = \textwidth]{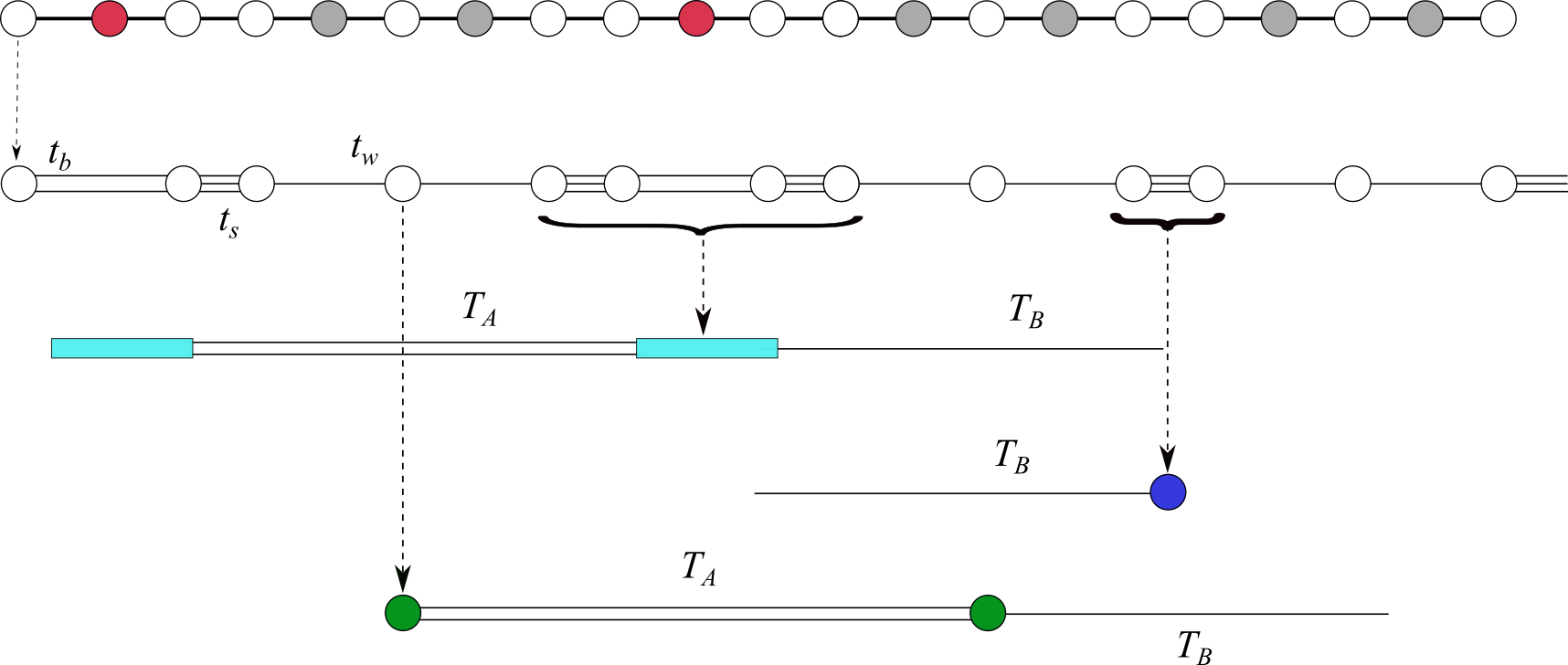}
		\caption{Decimation procedure for the effective chain described by $H_+$. For each of the three degenerate levels, we define a decimated chain with effective new couplings. This allows us to track how the different levels structure themselves. In this example, we focus on the chain resulting from the $v_A$ cluster. The couplings are first renormalized to three values $|t_s|>|t_b|>|t_w|$. The amount of lines connecting the unperturberd eigenstates are representative of the coupling strengths. At the next order of perturbation theory, the unperturbed  degenerate eigenstates are slightly more complicated, and correspond to four-atom molecules, dimers, and isolated sites. These then correspond to three chains with cyan, blue and green sites, respectively. The next corrections correspond to the Fibonacci case, where the levels trifurcate at each step. The hopping parameters $T_A$ and $T_B$ are calculated in \ref{App: Renormalized hoppings and energies.}.} 
		\label{fig:decimation}
	\end{figure}

\begin{figure*}[!bht]
		\centering
		\includegraphics[max width=\textwidth]{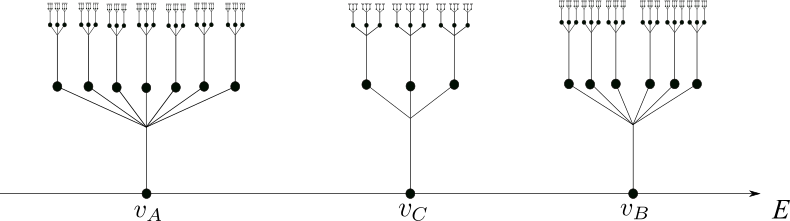}
		\caption{Energy splitting structure after each renormalization step. Energy scales are greatly distorted and do not represent actual gap sizes. Note that $v_C=v_B+2h$.}
		\label{fig:EnergySplitting}
	\end{figure*}

	We can apply the usual decimation procedure known for hierarchical chains \cite{NORI}. In this case, the novelty lies in the chain with three different on-site energies, as shown for example in \Cref{fig:decimation}. There are three renormalized chains that result from the first decimation step. We can analyze them separately to see how each energy level splits into different branches. A sketch of the branching structure is shown in \Cref{fig:EnergySplitting}, where one sees that there are three main clusters. The $v_C$ cluster follows the Fibonacci trifurcarting structure from the start. The $v_B$ cluster splits into six levels, each of which starts trifurcating according to the Fibonacci structure. Finally, the $v_A$ cluster splits into seven levels that also trifurcate afterwards. In the thermodynamic limit, these levels keep on trifurcating indefinitely, leading to a spectrum that is a Cantor set and known to be singular continuous \cite{Suto1989SingularHamiltonian}. 
    In \ref{App: Renormalized hoppings and energies.}, we provide a comprehensive explanation of the step-by-step renormalization process involved in handling the binary hoppings $T_A$ and $T_B$ for each cluster. In the next section, we show that these analytic tools lead to a good approximation and understanding of the structure of the spectrum. 
	
	\subsection{Effective Couplings and Energy Corrections}\label{EffectiveCalculations}

	In order to calculate the effective couplings and energy corrections at each splitting, we use the Brillouin-Wigner perturbation theory. For each cluster, we use the effective Hamiltonian \Cref{Eq: Effective Ham Deriv}
	\begin{equation*}\label{eq: Heff}
	H_{\rm{eff}}=QH_0Q+QH_1\sum_{n=0}^n\left(P\frac{1}{E-H_0}H_1\right)^nQ,
	\end{equation*}
	where $H_1$ is the perturbation corresponding to the weakest coupling, $Q$ is the projector onto the eigenspace of $H_0$ corresponding to the cluster of interest, while $P$ projects out of it. At first, $H_0$ is just composed of on-site energies and $H_1$ of the coupling $h$ between the isolated sites. This results in the three degenerate levels that form the main clusters. In the next order of perturbation theory, $H_0$ denotes the Hamiltonian of the corresponding renormalized chain, with the weakest coupling turned off. $H_1$ is then the perturbation with either $t_w$ or $t_b$ turned on, depending on which subchain we are dealing with (i.e. depending on whether $t_b>t_w$ or vice-versa).
    With this effective Hamiltonian for each of the clusters, the renormalized couplings can be found by calculating $\bra{E_j}H_{\rm{eff}}\ket{E_{j+1}}$, where $\ket{E_j}$ are the zeroth order local eigenstates, which are either one atomic site, a two or a four-atom molecule, depending on the situation. We now first state the results obtained from the first-order corrections, which we shall subsequently prove. We find the first three renormalized couplings to be given by 
\begin{empheq}[left = {\left(t_b,t_s,t_w\right)= \empheqlbrace}]{align}
		& \left( \frac{c}{1-2c},\ 1,\ c\right)h, \ \ \ \text{(Cluster A)} \notag \\
		& \left(-\frac{c^4}{2} ,\ -c ,\ c^2 \right)h, \ \ \ \text{(Cluster B)} \label{eq:tbtstw}
		\\
		& \left(0,\ -\frac{1}{4}\left[\frac{c}{1-2c}\right]^5,\  \frac{1}{16}\left[\frac{c}{1-2c}\right]^8 \right)h, \ \ \ \text{(Cluster C)} \notag 
	\end{empheq}
where we have defined the contrast parameter $c\equiv |h/(v_A-v_B)|$, which controls how well the perturbation theory behaves. We point out that the physics described by the theory is consistent for 
    \begin{equation}
        c<\frac{1}{4^{\frac{1}{3}}(1+4^{\frac{1}{6}})}\approx 0.27875,
    \end{equation}
    after which it fails to deliver reasonable results. This result will also be proven in the following paragraphs. 
    
\paragraph{First order renormalized hoppings}
	\begin{figure}
		\centering
		\includegraphics[width=\textwidth]{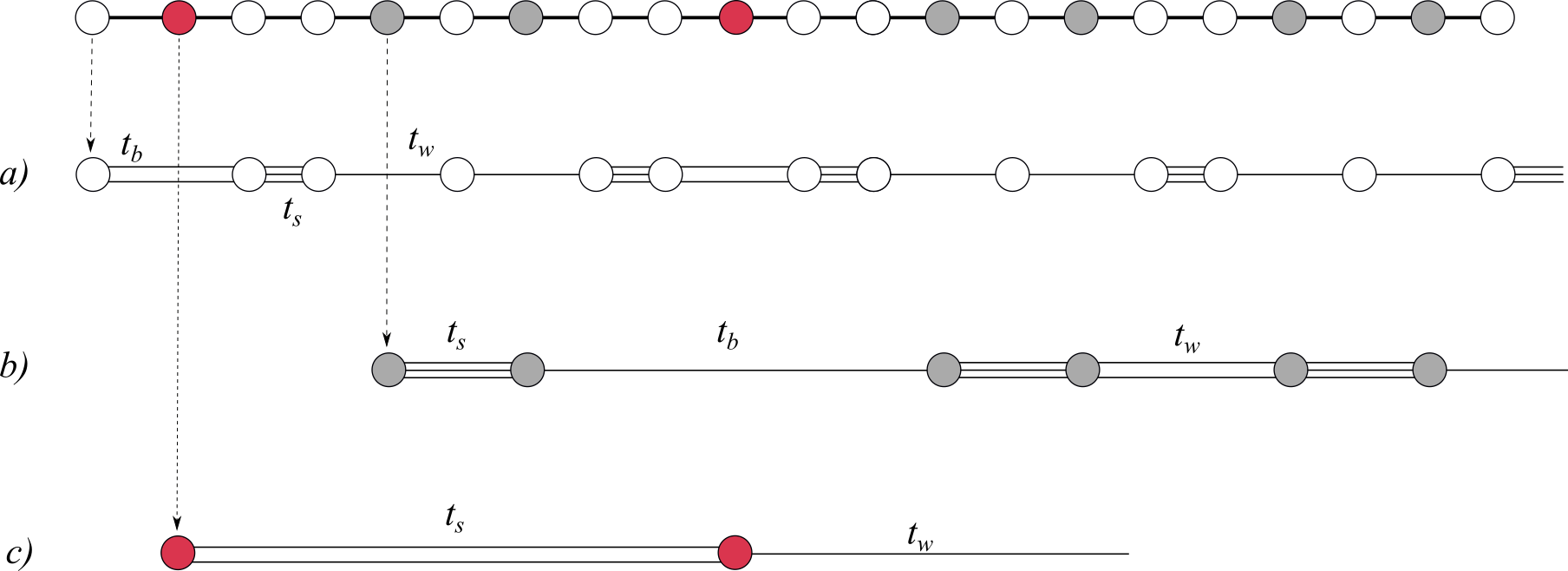}
		\caption{Decimation procedure. The topmost chain results from the equipartition theorem applied to the coupled Fibonacci chains we started with. It corresponds to the Hamiltonian $H_+$ with red sites of energies $v_C=v_B+2h$. The white and grey sites have energies $v_A$ and $v_B$ respectively. (a) The renormalized chain that was shown as an example in \Cref{sec:quasiPeriodicCoupling}, with $|t_s|>|t_b|>|t_w|$. (b) The chain corresponding to the renormalization of the grey sites. In this case, we have $|t_s|>|t_w|>|t_b|$. (c) The chain corresponding to the renormalization of the red sites, with $|t_s|>|t_w|$. This chain is already a proper hopping Fibonacci chain.
		}
		\label{fig:DecimationAll}
	\end{figure}
    \Cref{fig:DecimationAll} shows that one can form three effective chains (a), (b) and (c), corresponding to the clusters $v_A$, $v_B$ and $v_C=v_B+2h$ respectively. Before deriving our results, we recall that we are working in the regime of parameters where $v_A<v_B$ and $h<0$. This is done to obtain a hierarchy of hopping strengths that will result in the infinitely trifurcating structure after (at most) the second renormalization step. The $v_C$ cluster is a regular hopping Fibonacci chain. In fact, if one starts from a chain of generation $N$, with $F_N$ sites, one ends up with one of generation $N-5$, with $F_{N-5}$ sites. By calculating the matrix element $\bra{E_j}H_{\rm{eff}}\ket{E_{j+1}}$, where $H_{\rm{eff}}$ is given by \Cref{Eq: Effective Ham Deriv}, between the nearest neighbouring eigenstates of the unperturbed Hamiltonian $H_0$ in the $E=v_C$ subspace, we find that this chain has effective renormalized hoppings, to nearest order in the contrast $c=|h|/|v_A-v_B|$, given by
	\begin{equation}\label{eq:vchop}
	\begin{split}
	    t_s&=\frac{h^6}{4(v_B-v_A+2h)^5}=-\frac{1}{4}\left[\frac{c}{1-2c}\right]^5h, \\
	    t_w&=\frac{1}{16}\frac{h^9}{(v_B-v_A+2h)^8}=\frac{1}{16}\left[\frac{c}{1-2c}\right]^8h.
	\end{split}
	\end{equation}
	
	Next, we consider the $v_B$ chain, which contains six clusters. Two clusters correspond to the two-atom molecule bonded by $t_s$, while the other four correspond to the four-atom molecule (composed of two two-atom molecules bonded by $|t_w|>|t_b|$). These clusters will further start to trifurcate because they end up having the structure of a hopping Fibonacci chain after decimation. The renormalized hoppings of this chain are given by 
	\begin{equation}\label{eq:vbhop}
	\begin{split}
	     t_b&=\frac{-h^5}{2(v_B-v_A)^4}=-\frac{c^4}{2}h, \\ t_s&=\frac{h^2}{v_B-v_A}=-ch, \\
	     t_w&=\frac{h^3}{(v_B-v_A)^2}=c^2h. 
	\end{split}
	\end{equation}
	
	Finally, we consider the $v_A$ cluster. It has seven clusters. The six clusters come from the same structures as the $v_B$ chain, i.e two- and four-atoms molecules, and the additional cluster comes from isolated sites. Once again, each one of these clusters will start trifurcating at the next steps of decimation, and one retrieves the hopping Fibonacci chain structure. The renormalized hoppings of this chain are 
	\begin{equation}\label{vahop}
	\begin{split}
        t_b&=\frac{h^2}{v_A-v_B-2h}=\frac{c}{1-2c}h,\\
        t_s&=h, \\ 
        t_w&=\frac{h^2}{v_A-v_B}=ch.
	\end{split}
	\end{equation}
\begin{table}[!hbt]\label{table: energies}
    
            \begin{indented}
            \centering
               \lineup
        \item[]
		\begin{tabular}{|c|c|}
			\hline
			$E_0$  & $E_1$ 
			\\ \hline\hline
			\multirow{7}{*}{$v_A$} & $
			\frac{1}{\sqrt{2}}\sqrt{t_b^2+2t_s^2+\sqrt{(t_b^2+2t_s^2)^2-4t_s^4}}$ \\ \cline{2-2}
			& $
			\frac{1}{\sqrt{2}}\sqrt{t_b^2+2t_s^2-\sqrt{(t_b^2+2t_s^2)^2-4t_s^4}}$                    \\ \cline{2-2}
			& $t_s$                    \\ \cline{2-2}
			& 0                    \\ \cline{2-2}
			& $-t_s$                    \\ \cline{2-2}
			& $
			\frac{1}{\sqrt{2}}\sqrt{t_b^2+2t_s^2+\sqrt{(t_b^2+2t_s^2)^2-4t_s^4}}$                    \\ \cline{2-2}
			& $
			-\frac{1}{\sqrt{2}}\sqrt{t_b^2+2t_s^2-\sqrt{(t_b^2+2t_s^2)^2-4t_s^4}}$                    \\ \hline\hline
			\multirow{6}{*}{$v_B$} & $
			\frac{1}{\sqrt{2}}\sqrt{t_b^2+2t_s^2+\sqrt{(t_b^2+2t_s^2)^2-4t_s^4}}$ \\ \cline{2-2}
			& -$
			\frac{1}{\sqrt{2}}\sqrt{t_b^2+2t_s^2-\sqrt{(t_b^2+2t_s^2)^2-4t_s^4}}$                    \\ \cline{2-2}
			& $t_s$                    \\ \cline{2-2}
			& $-t_s$                    \\ \cline{2-2}
			& $
			-\frac{1}{\sqrt{2}}\sqrt{t_b^2+2t_s^2+\sqrt{(t_b^2+2t_s^2)^2-4t_s^4}}$                    \\ \cline{2-2}
			& $
			-\frac{1}{\sqrt{2}}\sqrt{t_b^2+2t_s^2-\sqrt{(t_b^2+2t_s^2)^2-4t_s^4}}$                   \\ \hline\hline
			\multirow{3}{*}{$v_B+2h$} & $t_s$ \\ \cline{2-2}
			& 0                    \\ \cline{2-2}
			& -$t_s$                    \\ \hline
		\end{tabular}
            \end{indented}
            \caption{The first order correction to the three clusters $v_A$, $v_B$ and $v_B+2h$. Any higher order correction will further split each of these energies into three, with corrections given by $\pm T_A$ and 0 (see \ref{App: Renormalized hoppings and energies.} for the values of $T_A$ in each case). Each of the three levels again will split into three and so on, with the well known spectrum structure of the hopping Fibonacci chain.}	
        \end{table}  
\paragraph{Regime of validity of the perturbation theory}	
	Let us now study the regime of validity which results in the hopping hierarchy discussed previously. Starting with the $v_B$ subspace, which is the simplest to deal with, we see from \Cref{eq:vbhop} that the hierarchy $|t_s|>|t_w|>|t_b|$ will always hold for $c\in(0,1)$. This immediately results in the trifurcating structure after one iteration of the renormalization procedure. The $v_A$ subspace, on the other hand, restricts the range of $c$ further. In order to have $|t_s|>|t_b|>|t_w|$, we must impose 
	\begin{equation*}
	    1>\left|\frac{c}{1-2c}\right|>c
	\end{equation*}
	which leads to $0<c<1/3$. Finally, the $v_C$ subspace will give us the final restriction to impose the hierarchy leading to a trifurcating structure. This means we want $|t_s|>|t_w|$, leading to
    \begin{equation*}
        \frac{1}{4}\left|\frac{c}{1-2c}\right|^5>\left|\frac{c}{1-2c}\right|^8 
    \end{equation*}
	Solving for $c>0$ yields the final restriction 
    \begin{equation}
        c<\frac{1}{4^{\frac{1}{3}}(1+4^{\frac{1}{6}})}.
    \end{equation}

\paragraph{Energy corrections}
    
\begin{figure}[!hbt]
		\centering
		\includegraphics[width = \textwidth]{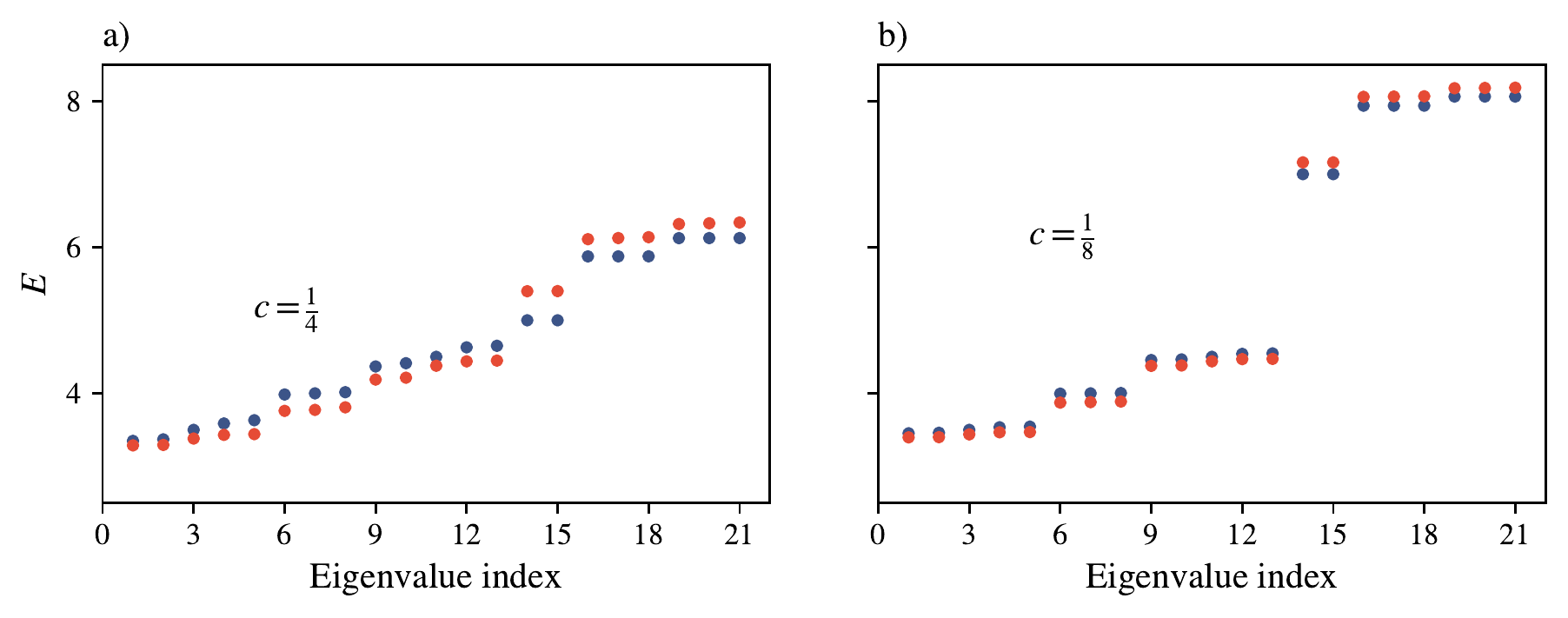}
		\caption{Comparison between the perturbative calculations (blue) and the direct numerical calculation (red) of the chain containing two ``impurities'' of strength $V_B+2h$, at the center of the resonator structure $AABAA$. The chain's length is 21 sites. The structure is the same and the discrepancy gets smaller as the contrast decreases. (a) Contrast $c=1/4$, which is close to the limits of applicability of the perturbation theory. (b) Contrast $c=1/8$. The energy is shown in arbitrary units.}
		\label{fig:comparisontheornumer}
	\end{figure}

    We also determined the first-order energy corrections using the same perturbation theory, and found the spectrum shown in  \Cref{fig:comparisontheornumer}. The energy levels calculated using perturbation theory (in blue) and those from numerical direct diagonalization (in red) have the same structure, with some discrepancies that disappear as the contrast $c\to0$. This is illustrated in \Cref{fig:comparisontheornumer} (a) and (b)  with $c=1/4$ and $c=1/8$, respectively. The structure of the spectrum is well approximated by the theory, even for a small chain size of 21 sites. The exact expressions for the energy levels can be found in Table 1. 
    	\begin{figure}[!hbt]
		\centering
		\includegraphics[scale=0.3]{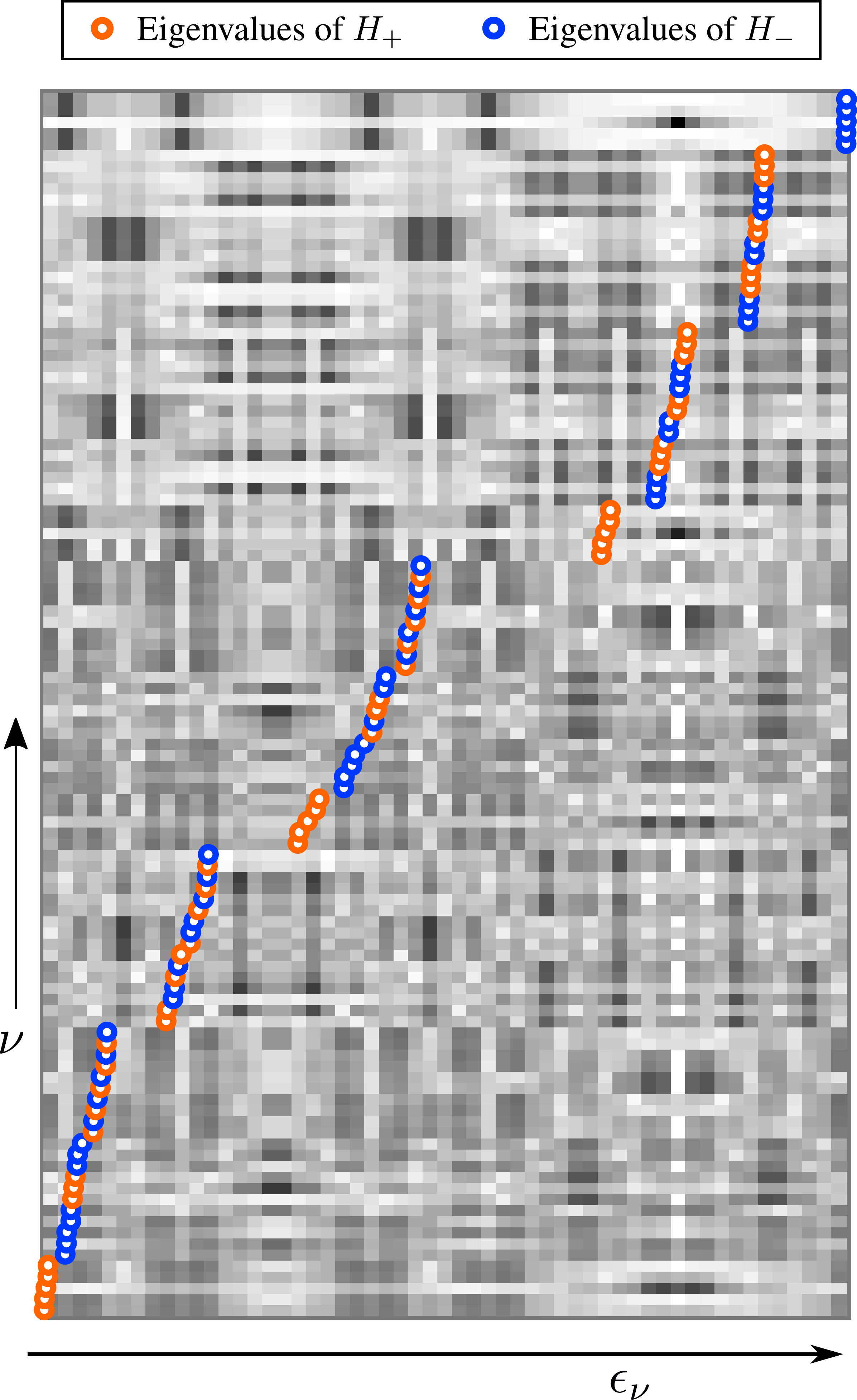}
		\caption{Eigenstate map for a setup consisting of two quasiperiodically coupled Fibonacci chains, each of length $N = 55$ sites.
		Orange dots correspond to eigenvalues of $H_{+}$ and blue dots correspond to eigenvalues of $H_{-}$. The color map is the same as in \cref{fig:isolatedRing}, with black pixels corresponding to $\max\limits_{\nu,j} \sqrt{|\Psi_j^{(\nu)}|}.$
		}
	    \label{fig:quasiPeriodicEigenstateMap}
	\end{figure}

    Now, one only needs to add the spectrum of the regular chain to this ``modified'' chain to obtain the full spectrum of the two connected chains, as shown in \Cref{fig:quasiPeriodicEigenstateMap}.

	\clearpage

	\section{Coupling through intermediate sites} \label{sec:IntermediateCoupling}
\begin{figure}[!hbt]
		\centering
		\includegraphics[max width = \textwidth]{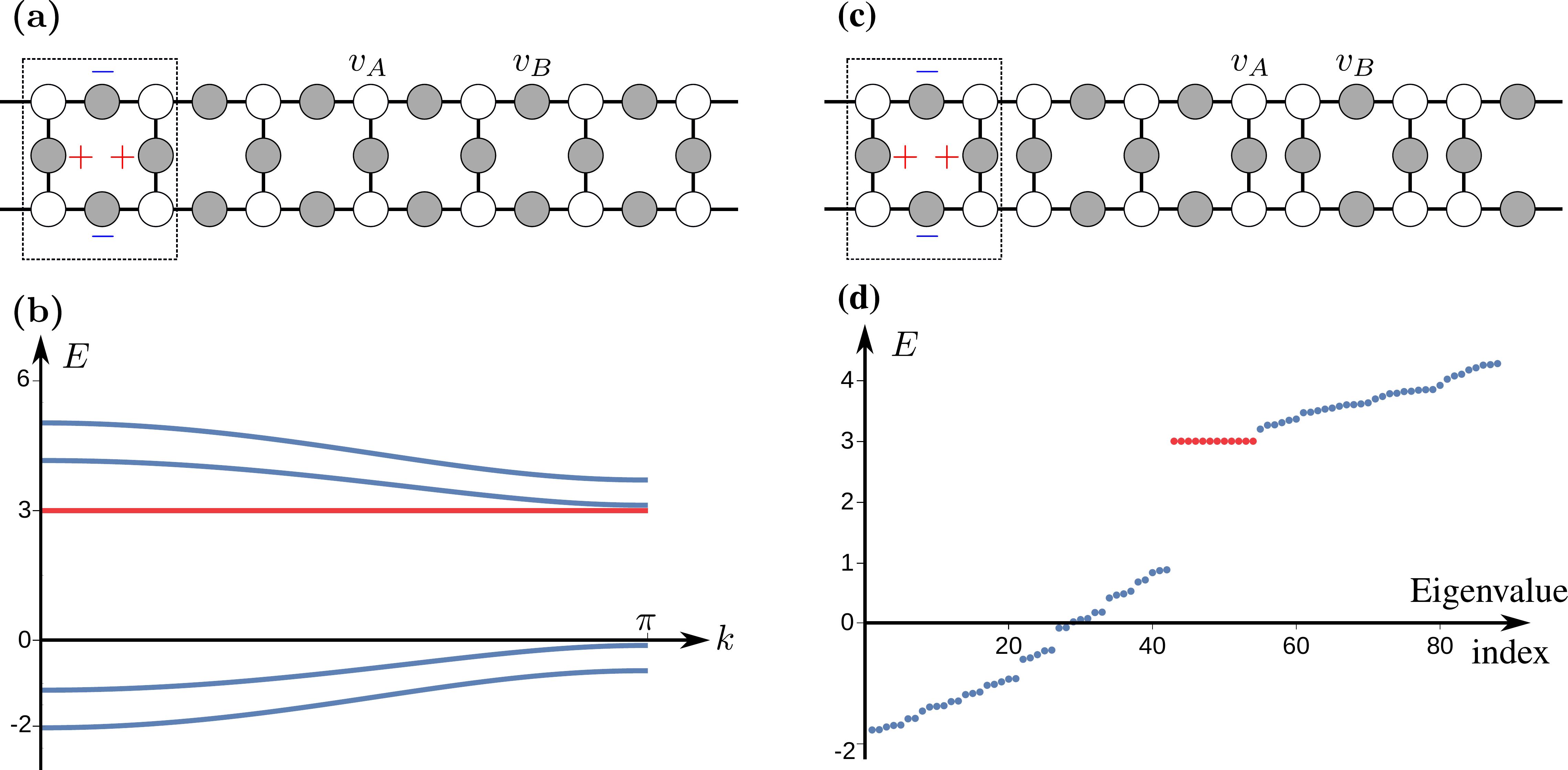}
		\caption{
        (a) The one-dimensional Lieb-lattice and its spectrum (b).
        (c) the quasiperiodic Lieb lattice and its spectrum (d).
		For the quasiperiodic Lieb lattice, the spectrum was computed for a finite setup with a total number of $88$ sites ($34$ in the upper and lower chain and $20$ $B$-sites in the center).
		For all figures, $v_A= 0$ and $v_B = h = 1$.
		}
		\label{fig:CLSs}
	\end{figure}

	In the following, we consider a case where the two Fibonacci chains are not coupled directly, but instead through intermediate sites. Before we start, let us first investigate a setup where two \emph{periodic} chains are coupled indirectly, as shown in \Cref{fig:CLSs} (a).
	In \Cref{fig:CLSs} (b), we depict the band structure of this so-called ``one-dimensional Lieb lattice'' \cite{Ramachandran2018FRiOaM219311FanoResonancesFlatBand}.
	What makes this lattice interesting is that, among four dispersive bands, it also features one completely flat band fulfilling $E(k) = \text{constant}$ for all $k$.
	This defining feature of flat bands renders them dispersionless; they suppress wave transport \cite{Leykam2018AP31473052ArtificialFlatBandSystems}.
	On the other hand, the density of states in a flat band diverges, so that any disorder or non-linear effects may qualitatively change the transport properties \cite{Leykam2018AP31473052ArtificialFlatBandSystems,Leykam2017EPJB90LocalizationWeaklyDisorderedFlat}.
	Moreover, in lattices where the single-particle Hamiltonian features a flat band, even arbitrarily weak interactions may act non-perturbatively.
	This can lead to boson pair formation \cite{Mielke2018JSP171679PairFormationHardCore,Pudleiner2015EPJB88207InteractingBosonsTwodimensionalFlat,Takayoshi2013PRA88063613PhaseDiagramPairTomonagaLuttinger} or other interesting phases, such as the Haldane insulator \cite{Gremaud2017PRB95165131HaldanePhaseSawtoothLattice} and Wigner crystals \cite{Tovmasyan2018148StronglyCorrelatedPhasesFlatband}.

	In two dimensions, a classical example for a system with flat bands is the Lieb lattice, whose structure is very similar to the system depicted in \Cref{fig:CLSs} (a).
	The two-dimensional Lieb lattice has been realized in a number of different experimental platforms, such as tailored atomic structures on substrates \cite{Slot2017NP13672ExperimentalRealizationCharacterizationElectronic,Drost2017NP13668TopologicalStatesEngineeredAtomic}, evanescently coupled waveguide arrays    \cite{Vicencio2015PRL114245503ObservationLocalizedStatesLieb,Mukherjee2015PRL114245504ObservationLocalizedFlatBandState}, terahertz spoof plasmons \cite{Kajiwara2016PRB93075126ObservationNonradiativeFlatBand}, or cold atom setups \cite{Taie2020NC11257SpatialAdiabaticPassageMassive,Taie2015SA1e1500854CoherentDrivingFreezingBosonic}.
    Interestingly, the CuO$_{2}$ planes in high-temperature cuprate superconductors possess a Lieb lattice structure, and it has been conjectured that flat bands might play a role in their high critical temperature \cite{Leykam2018AP31473052ArtificialFlatBandSystems,Peotta2015NC68944SuperfluidityTopologicallyNontrivialFlat,Julku2016PRL117045303GeometricOriginSuperfluidityLiebLattice,Kobayashi2016PRB94214501SuperconductivityRepulsivelyInteractingFermions,Tovmasyan2016PRB94245149EffectiveTheoryEmergentSU2,Liang2017PRB95024515BandGeometryBerryCurvature}.
	
	Flat bands are also interesting from another perspective, since they are tightly connected to the emergence of a special kind of eigenstates, the so-called ``compact localized state'' (CLS) \cite{Maimaiti2017PRB95115135CompactLocalizedStatesFlatband,Rhim2021APX61901606SingularFlatBands,Rhim2019PRB99045107ClassificationFlatBandsAccording}. For the one-dimensional Lieb lattice, a CLS is shown in \Cref{fig:CLSs} (a), consisting of an excitation of only four $B$ sites. 
	The CLS thus ``lives'' only on a single plaquette (marked by a dotted rectangle) and strictly vanishes outside it.
	In other words, it is perfectly localized on a very small part of the setup.
	
	The defining feature of CLSs --- namely, their perfect localization --- renders these states  very robust against perturbations: Since they vanish \emph{exactly} outside their localization domain $\mathcal{D}$, they are not affected by any changes to the system outside $\mathcal{D}$.
	Due to this property, CLSs are ideal candidates for storing information \cite{Rontgen2019PRL123080504QuantumNetworkTransferStorage,Kempkes2023QF21CompactLocalizedBoundaryStates}.
	The perfect localization of CLSs might further be interesting in the context of photonic waveguide arrays, where it allows for diffraction-free transmission of information in the form of CLSs \cite{Vicencio2013JO16015706DiffractionfreeImageTransmissionKagome,Vicencio2015PRL114245503ObservationLocalizedStatesLieb}.
	
	After the above considerations, let us now couple two aperiodic Fibonacci chains in an indirect manner.
	Out of the many possibilities, here we choose one of the simplest, yet quite interesting setup, depicted in \Cref{fig:CLSs} (c). Each $A$-site of the upper chain is coupled by an intermediate $B$-site to its counterpart $A$-site on the lower chain.
	This setup also features a macroscopic number of CLSs, one of which is depicted in \Cref{fig:CLSs} (c).

	
    When comparing the one-dimensional version of the Lieb lattice \cite{Lieb1989PRL621201TwoTheoremsHubbardModel,Flach2014E10530001DetanglingFlatBandsFano}---shown in \Cref{fig:CLSs} (a)---to the coupled Fibonacci chains depicted in \Cref{fig:CLSs} (c), one sees that they are rather similar.
    Due to this high similarity, we call the coupled Fibonacci setup of \Cref{fig:CLSs} (c) the ``quasiperiodic Lieb lattice''.
    The similarity in the structure of the two lattices is also visible in their eigenvalue spectra, since both lattices feature a flat (quasi) band.
    The emergence of this flat band can easily be understood, since (for an infinite setup) there is an infinite number of plaquettes, and thus an infinite number of degenerate CLSs.  In both cases, the flat bands emerge due to destructive interference \cite{Leykam2018AP31473052ArtificialFlatBandSystems}.

	
	\section{Conclusions} \label{sec:Conclusion}
	
	In this work, we analyzed various ways of coupling two identical Fibonacci chains to each other. The main tool that helped us to identify the features of these systems is a symmetry adapted block-diagonalization of the Hamiltonian $H$ into $H_+\oplus H_-$. Once we find the eigenstates of $H_\pm$, we can symmetrize/anti-symmetrize them, respectively, to obtain the eigenstates of the original Hamiltonian $H$. In addition, the eigenvalue spectrum of the system is simply given by a multiset sum of the eigenvalue spectra of the blocks $H_\pm$. 

	After briefly introducing the individual Fibonacci chain, we started by exploring the effects of uniformly coupling two identical chains. We then found that the resulting eigenvalue spectrum is just a sum of two shifted Fibonacci spectra, which renders the behaviour of a particle in such a system identical to that in a conventional Fibonacci chain. We have subsequently explored the case of a nonuniform coupling, where we coupled only $A$ (or only $B$) sites. An interesting scenario occurs when the interchain coupling is $h'=v_B-v_A$, since $H_+$ then becomes a periodic chain, for which the eigenstates are Bloch waves. On the other hand, $H_-$ is still a Fibonacci chain, such that the complete spectrum offers a mixture of critical and fully extended eigenstates, identifiable by the parity of the corresponding wavefunctions. The next type of coupling that we have analyzed is between a small subset of sites, i.e. a so called defect coupling, leading to block Hamiltonians $H_\pm$ which are Fibonacci chains with on-site defects. This has been followed by another interesting and more complicated case of quasiperiodically coupled chains. The resulting block Hamiltonians could be thought to be like several coupled defects, but we have shown, through a perturbative renormalization analysis, that all states in this chain belong to the same family of critical states. Finally, we explored two Fibonacci chains coupled to each other in the same manner as the nonuniform coupling of \Cref{sec:non-UniformCoupling}, but with an intermediate site in between. This offers the possibility of having a set of CLSs, leading to a flat bland in the energy spectrum. Overall, the demonstrated emergence of CLSs in the quasiperiodic Lieb lattice represents an interesting addition to the existing literature on these phenomena in quasiperiodic setups (see, for instance, \cite{Sutherland1986PRB345208LocalizationElectronicWaveFunctions,Ha2021PRB104165112MacroscopicallyDegenerateLocalizedZeroenergy,Kohmoto1986PRB343849ElectronicVibrationalModesPenrose}).
    Two immediate tasks for the near future would be to analyze the quasiperiodic Lieb lattice in the context of interacting electrons, or to  investigate the impact of (correlated) disorder on the transport properties.
    

	\section{Acknowledgments}
    This publication is part of the project TOPCORE with project number OCENW.GROOT.2019.048, which is financed by the Dutch Research Council (NWO). This work (P.S.) is supported by the Cluster of Excellence 'Advanced Imaging of Matter' of the Deutsche  Forschungsgemeinschaft (DFG) - EXC 2056 - project ID 390715994.

	* A.M. and M.R. contributed equally to this work.
\newline \newline
	\bibliography{Bibtexnew,MAINnew,MAIN2new,referencesnew} 

\providecommand{\noopsort}[1]{}\providecommand{\singleletter}[1]{#1}%
\providecommand{\newblock}{}
\begin{thebibliography}{10}
\expandafter\ifx\csname url\endcsname\relax
  \def\url#1{{\tt #1}}\fi
\expandafter\ifx\csname urlprefix\endcsname\relax\def\urlprefix{URL }\fi
\providecommand{\eprint}[2][]{\url{#2}}

\bibitem{shechtman}
Shechtman D, Blech I, Gratias D and Cahn J~W 1984 {\em Phys. Rev. Lett.\/} {\bf
  53}(20) 1951--1953
  \urlprefix\url{https://link.aps.org/doi/10.1103/PhysRevLett.53.1951}

\bibitem{Janssen:a25379}
Janssen T 1986 {\em Act. Crystall. Sect. A\/} {\bf 42} 261--271

\bibitem{alpdreelec}
Berger C, Grenet T, Lindqvist P, Lanco P, Grieco J, Fourcaudot G and
  Cyrot-Lackmann F 1993 {\em Solid State Communications\/} {\bf 87} 977 -- 979
  ISSN 0038-1098
  \urlprefix\url{http://www.sciencedirect.com/science/article/pii/003810989390543V}

\bibitem{Vieira2005Low-EnergyChains}
Vieira A~P 2005 {\em Physical Review Letters\/} {\bf 94} 077201 ISSN 0031-9007
  \urlprefix\url{https://link.aps.org/doi/10.1103/PhysRevLett.94.077201}

\bibitem{Tanese2014FractalPotential}
Tanese D, Gurevich E, Baboux F, Jacqmin T, Lema{\^{i}}tre A, Galopin E, Sagnes
  I, Amo A, Bloch J and Akkermans E 2014 {\em Physical Review Letters\/} {\bf
  112} 146404 ISSN 0031-9007
  \urlprefix\url{https://link.aps.org/doi/10.1103/PhysRevLett.112.146404}

\bibitem{Jagannathan2021RMP93045001FibonacciQuasicrystalCaseStudy}
Jagannathan A 2021 {\em Rev. Mod. Phys.\/} {\bf 93} 045001

\bibitem{emaciaaporder}
Maciá E 2005 {\em Reports on Progress in Physics\/} {\bf 69} 397--441
  \urlprefix\url{https://doi.org/10.1088%2F0034-4885%2F69%2F2%2Fr03}

\bibitem{tedjanssen}
de~Boissieu M 2019 {\em Act. Crystall. Sect. A\/} {\bf 75(Pt 2)} 273–280.

\bibitem{Niu1986PRL572057RenormalizationGroupStudyOneDimensionalQuasiperiodic}
Niu Q and Nori F 1986 {\em Phys. Rev. Lett.\/} {\bf 57} 2057--2060 ISSN
  0031-9007

\bibitem{Mac__2016}
Mac\'e N, Jagannathan A and Pi\'echon F 2016 {\em Phys. Rev. B\/} {\bf 93}(20)
  205153 \urlprefix\url{https://link.aps.org/doi/10.1103/PhysRevB.93.205153}

\bibitem{Moreira2006SpecificSequences}
Moreira D~A, Albuquerque E~L and Bezerra C~G 2006 {\em The European Physical
  Journal B\/} {\bf 54} 393--398 ISSN 1434-6028

\bibitem{Pal2014AbsolutelyNetworks}
Pal B and Chakrabarti A 2014 {\em Physica E: Low-dimensional Systems and
  Nanostructures\/} {\bf 60} 188--195 ISSN 1386-9477

\bibitem{Mukherjee2017EPJB9052ControlledDelocalizationElectronicStates}
Mukherjee A, Nandy A and Chakrabarti A 2017 {\em Eur. Phys. J. B\/} {\bf 90} 52
  ISSN 1434-6036

\bibitem{Saha2019ParticleLadder}
Saha M and Maiti S~K 2019 {\em Journal of Physics D: Applied Physics\/} {\bf
  52} 465304 ISSN 0022-3727
  \urlprefix\url{https://iopscience.iop.org/article/10.1088/1361-6463/ab3a0e}

\bibitem{Roy2022LocalizationEdge}
Roy S, Maiti S~K, P{\'{e}}rez L~M, Silva J~H~O and Laroze D 2022 {\em
  Materials\/} {\bf 15} 597 ISSN 1996-1944

\bibitem{KILIC2008701}
Kilic E 2008 {\em European Journal of Combinatorics\/} {\bf 29} 701--711 ISSN
  0195-6698
  \urlprefix\url{https://www.sciencedirect.com/science/article/pii/S0195669807000595}

\bibitem{SireMosseri}
Sire C and Mosseri R 1990 {\em Journal de Physique\/} {\bf 51} 1569 -- 1583

\bibitem{NORI}
Niu Q and Nori F 1991 {\em Phys. Rev. B\/} {\bf 42} 10329--10341

\bibitem{Kohmoto1986PRB34563QuasiperiodicLatticeElectronicProperties}
Kohmoto M and Banavar J~R 1986 {\em Physical Review B\/} {\bf 34} 563--566 ISSN
  0163-1829

\bibitem{Kohmoto1987PRB351020CriticalWaveFunctionsCantorset}
Kohmoto M, Sutherland B and Tang C 1987 {\em Phys. Rev. B\/} {\bf 35}
  1020--1033

\bibitem{Kohmoto1983PRL501870LocalizationProblemOneDimension}
Kohmoto M, Kadanoff L~P and Tang C 1983 {\em Phys. Rev. Lett.\/} {\bf 50}
  1870--1872

\bibitem{Rontgen2019PRB99214201LocalSymmetryTheoryResonator}
R{\"o}ntgen M, Morfonios C~V, Wang R, Dal~Negro L and Schmelcher P 2019 {\em
  Phys. Rev. B\/} {\bf 99} 214201

\bibitem{Szameit2012DiscreteOpticsFemtosecondLaser}
Szameit A, Dreisow F and Nolte S 2012 Discrete optics in femtosecond laser
  written waveguide arrays {\em Femtosecond Laser Micromachining: Photonic and
  Microfluidic Devices in Transparent Materials\/} ({\em Topics in {{Applied
  Physics}}\/} no 123) ({Springer, Berlin, Heidelberg}) pp 351--388 ISBN
  978-3-642-23365-4

\bibitem{Lee2018CP11TopolectricalCircuits}
Lee C~H, Imhof S, Berger C, Bayer F, Brehm J, Molenkamp L~W, Kiessling T and
  Thomale R 2018 {\em Commun Phys\/} {\bf 1} 1--9 ISSN 2399-3650

\bibitem{Dong2021PRR3023056TopolectricCircuitsTheoryConstruction}
Dong J, Juri{\v c}i{\'c} V and Roy B 2021 {\em Phys. Rev. Research\/} {\bf 3}
  023056

\bibitem{Lazo2000IaNSV387MultifractalBehaviorFibonacciCrystal}
Lazo E 2000 Multifractal {{Behavior}} of a {{Fibonacci Crystal Built}} over p
  {{Coupled Chains}} {\em Instabilities and {{Nonequilibrium Structures VI}}\/}
  Nonlinear {{Phenomena}} and {{Complex Systems}} ed Tirapegui E, Mart{\'i}nez
  J and Tiemann R ({Dordrecht}: {Springer Netherlands}) pp 387--392 ISBN
  978-94-011-4247-2

\bibitem{archambault_janot_1997}
Archambault P and Janot C 1997 {\em MRS Bulletin\/} {\bf 22} 48–53

\bibitem{refId0}
{Janot, C} 1996 {\em Europhys. News\/} {\bf 27} 60--64
  \urlprefix\url{https://doi.org/10.1051/epn/19962702060}

\bibitem{subsdisorder}
Naumis G~G and Arag\'on J~L 1996 {\em Phys. Rev. B\/} {\bf 54}(21) 15079--15085
  \urlprefix\url{https://link.aps.org/doi/10.1103/PhysRevB.54.15079}

\bibitem{Niu1990PRB4210329SpectralSplittingWavefunctionScaling}
Niu Q and Nori F 1990 {\em Physical Review B\/} {\bf 42} 10329--10341 ISSN
  0163-1829, 1095-3795

\bibitem{Moustaj2021PRB104144201EffectsDisorderFibonacciQuasicrystal}
Moustaj A, Kempkes S and Smith C~M 2021 {\em Phys. Rev. B\/} {\bf 104} 144201

\bibitem{Suto1989SingularHamiltonian}
S{\"{u}}t{\H{o}} A 1989 {\em Journal of Statistical Physics\/} {\bf 56}
  525--531 ISSN 0022-4715

\bibitem{Ramachandran2018FRiOaM219311FanoResonancesFlatBand}
Ramachandran A, Danieli C and Flach S 2018 Fano resonances in flat band
  networks {\em Fano {{Resonances}} in {{Optics}} and {{Microwaves}}\/} vol 219
  ed Kamenetskii E, Sadreev A and Miroshnichenko A ({Cham}: {Springer
  International Publishing}) pp 311--329 ISBN 978-3-319-99730-8
  978-3-319-99731-5

\bibitem{Leykam2018AP31473052ArtificialFlatBandSystems}
Leykam D, Andreanov A and Flach S 2018 {\em Adv. Phys.\/} {\bf 3} 1473052 ISSN
  null

\bibitem{Leykam2017EPJB90LocalizationWeaklyDisorderedFlat}
Leykam D, Bodyfelt J~D, Desyatnikov A~S and Flach S 2017 {\em The European
  Physical Journal B\/} {\bf 90} 1 ISSN 1434-6028, 1434-6036

\bibitem{Mielke2018JSP171679PairFormationHardCore}
Mielke A 2018 {\em J Stat Phys\/} {\bf 171} 679--695 ISSN 1572-9613

\bibitem{Pudleiner2015EPJB88207InteractingBosonsTwodimensionalFlat}
Pudleiner P and Mielke A 2015 {\em Eur. Phys. J. B\/} {\bf 88} 207 ISSN
  1434-6036

\bibitem{Takayoshi2013PRA88063613PhaseDiagramPairTomonagaLuttinger}
Takayoshi S, Katsura H, Watanabe N and Aoki H 2013 {\em Phys. Rev. A\/} {\bf
  88} 063613

\bibitem{Gremaud2017PRB95165131HaldanePhaseSawtoothLattice}
Gr{\'e}maud B and Batrouni G~G 2017 {\em Phys. Rev. B\/} {\bf 95} 165131

\bibitem{Tovmasyan2018148StronglyCorrelatedPhasesFlatband}
Tovmasyan M 2018 {\em Strongly Correlated Phases in Flatband Lattices\/} Ph.D.
  thesis ETH Zurich

\bibitem{Slot2017NP13672ExperimentalRealizationCharacterizationElectronic}
Slot M~R, Gardenier T~S, Jacobse P~H, {van Miert} G~C~P, Kempkes S~N,
  Zevenhuizen S~J~M, Smith C~M, Vanmaekelbergh D and Swart I 2017 {\em Nat.
  Phys.\/} {\bf 13} 672--676 ISSN 1745-2473, 1745-2481

\bibitem{Drost2017NP13668TopologicalStatesEngineeredAtomic}
Drost R, Ojanen T, Harju A and Liljeroth P 2017 {\em Nat. Phys.\/} {\bf 13}
  668--671 ISSN 1745-2481

\bibitem{Vicencio2015PRL114245503ObservationLocalizedStatesLieb}
Vicencio R~A, Cantillano C, {Morales-Inostroza} L, Real B,
  {Mej{\'i}a-Cort{\'e}s} C, Weimann S, Szameit A and Molina M~I 2015 {\em Phys.
  Rev. Lett.\/} {\bf 114} 245503

\bibitem{Mukherjee2015PRL114245504ObservationLocalizedFlatBandState}
Mukherjee S, Spracklen A, Choudhury D, Goldman N, {\"O}hberg P, Andersson E and
  Thomson R~R 2015 {\em Phys. Rev. Lett.\/} {\bf 114} 245504

\bibitem{Kajiwara2016PRB93075126ObservationNonradiativeFlatBand}
Kajiwara S, Urade Y, Nakata Y, Nakanishi T and Kitano M 2016 {\em Phys. Rev.
  B\/} {\bf 93} 075126

\bibitem{Taie2020NC11257SpatialAdiabaticPassageMassive}
Taie S, Ichinose T, Ozawa H and Takahashi Y 2020 {\em Nat. Commun.\/} {\bf 11}
  257 ISSN 2041-1723

\bibitem{Taie2015SA1e1500854CoherentDrivingFreezingBosonic}
Taie S, Ozawa H, Ichinose T, Nishio T, Nakajima S and Takahashi Y 2015 {\em
  Sci. Adv.\/} {\bf 1} e1500854 ISSN 2375-2548

\bibitem{Peotta2015NC68944SuperfluidityTopologicallyNontrivialFlat}
Peotta S and T{\"o}rm{\"a} P 2015 {\em Nat. Commun.\/} {\bf 6} 8944 ISSN
  2041-1723

\bibitem{Julku2016PRL117045303GeometricOriginSuperfluidityLiebLattice}
Julku A, Peotta S, Vanhala T~I, Kim D~H and T{\"o}rm{\"a} P 2016 {\em Phys.
  Rev. Lett.\/} {\bf 117} 045303

\bibitem{Kobayashi2016PRB94214501SuperconductivityRepulsivelyInteractingFermions}
Kobayashi K, Okumura M, Yamada S, Machida M and Aoki H 2016 {\em Phys. Rev.
  B\/} {\bf 94} 214501

\bibitem{Tovmasyan2016PRB94245149EffectiveTheoryEmergentSU2}
Tovmasyan M, Peotta S, T{\"o}rm{\"a} P and Huber S~D 2016 {\em Phys. Rev. B\/}
  {\bf 94} 245149

\bibitem{Liang2017PRB95024515BandGeometryBerryCurvature}
Liang L, Vanhala T~I, Peotta S, Siro T, Harju A and T{\"o}rm{\"a} P 2017 {\em
  Phys. Rev. B\/} {\bf 95} 024515

\bibitem{Maimaiti2017PRB95115135CompactLocalizedStatesFlatband}
Maimaiti W, Andreanov A, Park H~C, Gendelman O and Flach S 2017 {\em Phys. Rev.
  B\/} {\bf 95} 115135

\bibitem{Rhim2021APX61901606SingularFlatBands}
Rhim J~W and Yang B~J 2021 {\em Advances in Physics: X\/} {\bf 6} 1901606 ISSN
  null

\bibitem{Rhim2019PRB99045107ClassificationFlatBandsAccording}
Rhim J~W and Yang B~J 2019 {\em Phys. Rev. B\/} {\bf 99} 045107 ISSN 2469-9950,
  2469-9969

\bibitem{Rontgen2019PRL123080504QuantumNetworkTransferStorage}
R{\"o}ntgen M, Morfonios C~V, Brouzos I, Diakonos F~K and Schmelcher P 2019
  {\em Phys. Rev. Lett.\/} {\bf 123} 080504

\bibitem{Kempkes2023QF21CompactLocalizedBoundaryStates}
Kempkes S~N, Capiod P, Ismaili S, Mulkens J, Eek L, Swart I and Morais~Smith C
  2023 {\em Quantum Front\/} {\bf 2} 1 ISSN 2731-6106

\bibitem{Vicencio2013JO16015706DiffractionfreeImageTransmissionKagome}
Vicencio R~A and {Mej{\'i}a-Cort{\'e}s} C 2013 {\em J. Opt.\/} {\bf 16} 015706
  ISSN 2040-8986

\bibitem{Lieb1989PRL621201TwoTheoremsHubbardModel}
Lieb E~H 1989 {\em Phys. Rev. Lett.\/} {\bf 62} 1201--1204

\bibitem{Flach2014E10530001DetanglingFlatBandsFano}
Flach S, Leykam D, Bodyfelt J~D, Matthies P and Desyatnikov A~S 2014 {\em
  Europhys. Lett.\/} {\bf 105} 30001 ISSN 0295-5075

\bibitem{Sutherland1986PRB345208LocalizationElectronicWaveFunctions}
Sutherland B 1986 {\em Phys. Rev. B\/} {\bf 34} 5208--5211

\bibitem{Ha2021PRB104165112MacroscopicallyDegenerateLocalizedZeroenergy}
Ha H and Yang B~J 2021 {\em Phys. Rev. B\/} {\bf 104} 165112 ISSN 2469-9950,
  2469-9969

\bibitem{Kohmoto1986PRB343849ElectronicVibrationalModesPenrose}
Kohmoto M and Sutherland B 1986 {\em Phys. Rev. B\/} {\bf 34} 3849--3853

\end{thebibliography}
\clearpage

\appendix	
	\section{Determination of effective hopping amplitudes and energy corrections of the quasiperiodic coupling case}\label{App: Renormalized hoppings and energies.}

    In the next subsections, we will determine the energy corrections and the renormalized Fibonacci hoppings for each cluster.

	\subsection{$v_A$ cluster}
	\begin{figure}[!hbt]
		\centering
		\includegraphics[width=\textwidth]{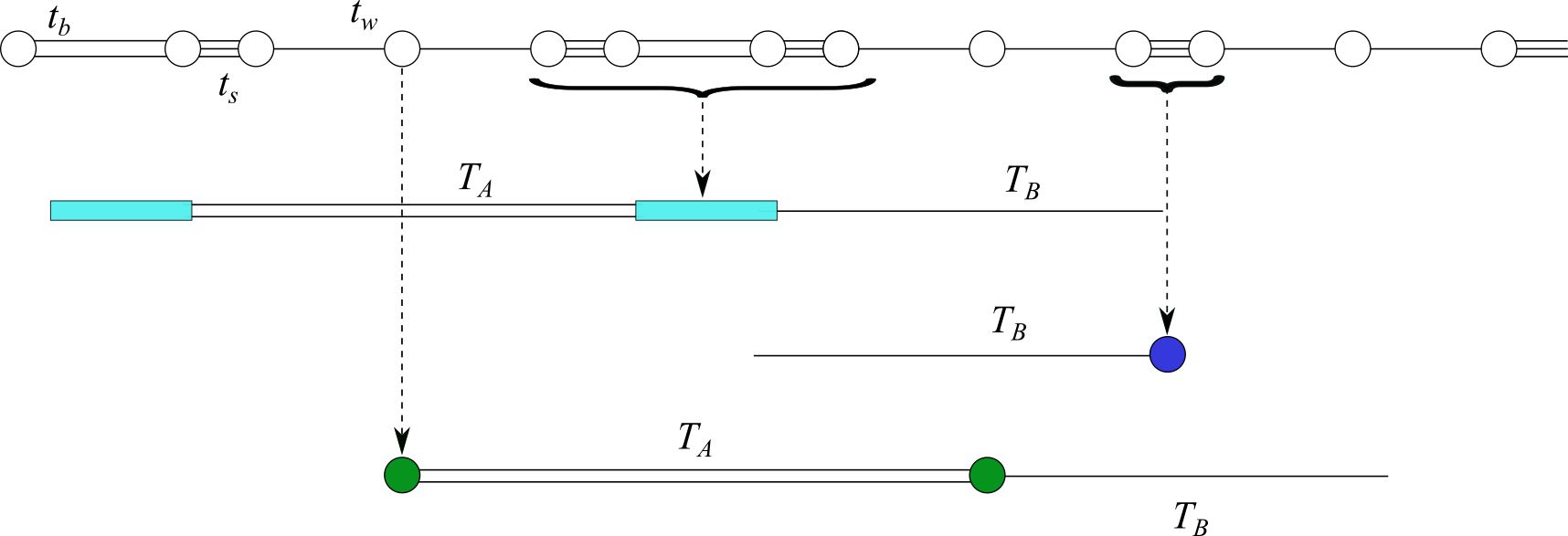}
		\caption{Decimation of the $v_A$ cluster's effective chain. The cyan rectangles represent the states corresponding to the four-atom molecules. They correspond to the renormalized sites of the chain, and have an on-site energy given by $E_{4M}$. The effective hoppings $T_A$ and $T_B$ are Fibonacci distributed and further renormalization leads to the Fibonacci trifurcating structure. Below, we have the two-atom molecular chain, where each blue site represents a dimer. In this figure, only one of them appears, but for a larger chain, there are Fibonacci distributed couplings, whence the trifurcating structure starts after this point in the renormalization procedure. Finally, the bottom chain corresponds to isolated (green) sites, forming the last degenerate subspace of $H_0^A$. The effective chain for this subspace also has couplings that are Fibonacci distributed.}
		\label{fig:DecimationGREEN}
	\end{figure}
	Starting with the $v_A$ cluster, we can split it into the three categories depicted in \Cref{fig:DecimationGREEN}. The 4-atom molecule leads to four energy eigenvalues, 
	\begin{equation}
	E^A_{4M}=\pm\frac{1}{\sqrt{2}}\sqrt{t_b^2+2t_s^2\pm\sqrt{(t_b^2+2t_s^2)^2-4t_s^4}},
	\end{equation}
	where the subscript $4M$ reminds us that it is for the 4-atom molecule, while the superscript $A$ indicates that it applies to the initial level $v_A$. The 2-atom molecule and the isolated atom have eigenvalues, respectively, equal to
	\begin{align}
	E^A_{2M}&=\pm t_s, \\
	E^A_A&=0.
	\end{align}
	
	In order to calculate the effective hoppings using the Brillouin-Wigner degenerate perturbation theory, we need to know the eigenstates of $H^A_0$, which for the $E_A^A=0$ level are just the corresponding isolated sites. For the 2-atom molecules, these are given by $\ket{E^A_{2M}}=2^{-1/2}(\ket{i}\pm\ket{i+1})$ for some localized one-particle state at site number $i$, corresponding to the sites coupled by the strongest bond, and for the 4-atom molecules
	\begin{equation*}
	\ket{E^A_{4M}}=\frac{a_0\ket{i}+a_1\ket{i+1}+a_2\ket{i+2}+a_3\ket{i+3}}{\sqrt{a^2_0+a_1^2+a_2^2+a_3^2}}
	\end{equation*}
	for the relevant sites $i$ to $i+3$, as shown in \Cref{fig:DecimationGREEN}. The coefficients can be computed by solving the eigenvalue problem 
 \begin{equation*}
     \begin{pmatrix}
         0 & t_s & 0 & 0 \\
         t_s & 0 & t_b & 0 \\
         0 & t_b & 0 & t_s \\
         0 & 0 & t_s & 0
     \end{pmatrix}\begin{pmatrix}
         a_0 \\ a_1 \\ a_2 \\ a_3
     \end{pmatrix}=E_{4M}\begin{pmatrix}
         a_0 \\ a_1 \\ a_2 \\ a_3
     \end{pmatrix}.
 \end{equation*}
 A solution to this problem is 
	\begin{equation}
	\begin{aligned}
	a_0&=1, &
	a_1&=\frac{E_{4M}}{t_s}, \\
	a_2&=\frac{E^2_{4M}-t_s^2}{t_bt_s}, &
	a_3&=\frac{E_{4M}}{t_s}\frac{E^2_{4M}-t_s^2-t_b^2}{t_bt_s},
	\end{aligned}
	\end{equation}
    where we ommited the superscript $A$ for brevity. Each of the degenerate subspaces can be associated to its own chain, with its own couplings in the renormalization picture (see \Cref{fig:DecimationGREEN}). 
	
	The effective Hamiltonian for the $v_A$ cluster is found by setting the perturbation Hamiltonian $H_1$ to be the one with all matrix elements containing $t_w$. We remind the reader that the general expression was given by \Cref{eq: Heff}.
	\paragraph{$\boldsymbol{E_{4M}}$ Chain (cyan)} The couplings in the 4-atom molecular chain are given by the matrix element $\bra{E^{A}_{4M,j}}H^{A}_{eff}\ket{E^{A}_{4M,j+1}}$, where $j$ is the renormalized site index. To nearest order in $t_w$, they read 
	\begin{equation}
	\begin{aligned}
	T_A&=\frac{a_3}{N}\frac{t_w^2}{E_{4M}}, \\
	T_B&= \frac{a_3}{N}\frac{t_w^4t_s}{E^2_{4M}(E^2_{4M}-t_s^2)},
	\end{aligned}
	\end{equation}
	where we defined $N\equiv a_0^2+a_1^2+a_2^2+a_3^2$. Note that there are four different $E_{4M}$'s and hence four sets of coefficients $a$.
	\paragraph{$\boldsymbol{E_{2M}}$ Chain (blue)} The 2-atom molecular chain's couplings are obtained from the matrix elements $\bra{E^{A}_{2M,j}}H^{A}_{eff}\ket{E^{A}_{2M,j+1}}$. The calculation yields
	\begin{equation}
	\begin{aligned}
	T_A&=\pm\frac{t_w^4}{2t_s^2}\sum_{j=0}^3\frac{a_3^{(j)}}{N_j\left(\pm t_s-E^{(j)}_{4M}\right)}, \\
	T_B&= \frac{t_w^6}{2t_s^3}\left(\sum_{j=0}^3\frac{a_3^{(j)}}{N_j\left(\pm t_s-E^{(j)}_{4M}\right)}\right)^2,
	\end{aligned}
	\end{equation}
	where we have now explicitely given a label to each of the four sets of $a_j$'s, and by extention, $N_j$'s as well. The $\pm$ refers to the bonding and anti-bonding energy levels.
	\paragraph{$\boldsymbol{E_{A}}$ Chain (green)} Finally, the isolated atom chain has the following couplings 
	\begin{equation}
	\begin{aligned}
	T_A&=-\frac{t_w^2}{2t_s}, \\
	T_B&=-t_w^2 \sum_{j=0}^3\frac{a_3^{(j)}}{N_jE^{(j)}_{4M}}.
	\end{aligned}
	\end{equation}
	\subsection{$v_B$ cluster}
	\begin{figure}[!t]
		\centering
		\includegraphics[width=\textwidth]{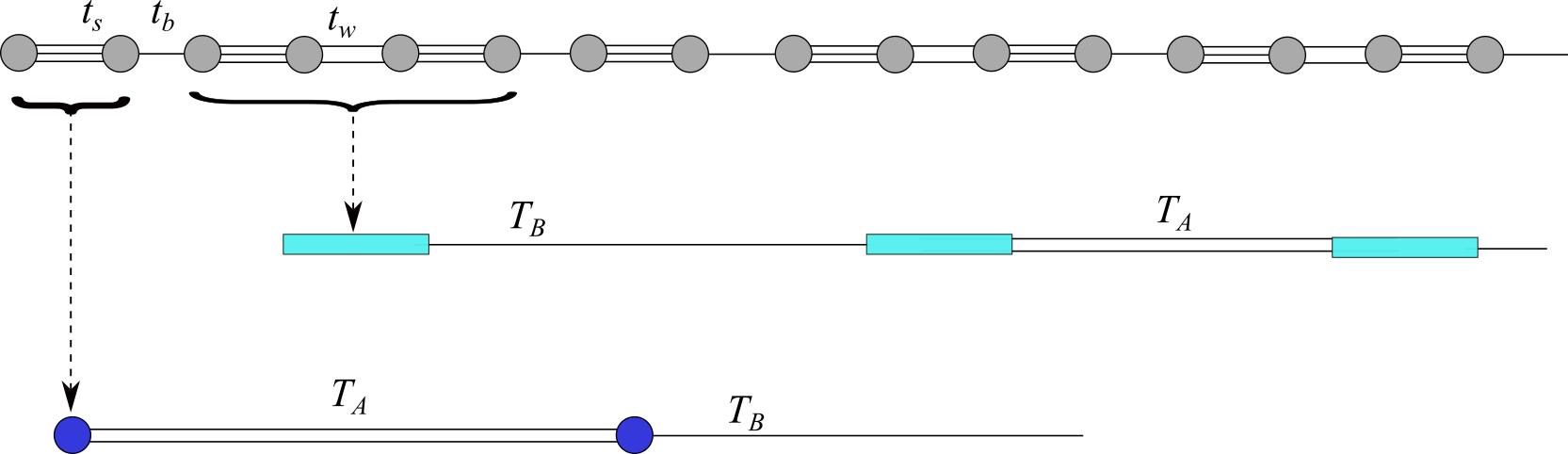}
		\caption{Decimation of the $v_B$ cluster's effective chain. The cyan rectangles represent the states corresponding to the four-atom molecules, like in \Cref{fig:DecimationGREEN}. Below, we have the blue sites representing dimers, also similar to \Cref{fig:DecimationGREEN}. The couplings in both cases follow a Fibonacci sequence and will also trifurcate after this step in the renormalization procedure.}
		\label{fig:DecimationRed}
	\end{figure}
	For the $v_B$ cluster's effective chains, we only have two categories, as shown in \Cref{fig:DecimationRed}. The first (cyan) corresponds to a 4-atom molecular chain, with the same energy eigenvalues and eigenstates formulae as the previous ones, but with a different set of $\{t_b,t_s,t_w\}$ [see \Cref{eq:tbtstw}]. The second (blue) subchain corresponds to the 2-atom molecular chain. We shall, once again, refer to them as the $E_{4M}$ and $E_{2M}$ chains, respectively.
	\paragraph{$\boldsymbol{E_{4M}}$ Chain} The couplings in this case are given by 
	\begin{equation}
	\begin{aligned}
	T_A&=\frac{t_w}{N}a_3, \\
	T_B&=\frac{a_3}{N}\frac{t_w^2t_s}{E_{4M}^2-t_s^2}.
	\end{aligned}
	\end{equation}
	
	\paragraph{$\boldsymbol{E_{2M}}$ Chain} The couplings for this chain are 
	\begin{equation}
	\begin{aligned}
	T_A&=\pm\frac{t_w^2}{2}\sum_{j=0}^3\frac{a_3^{(j)}}{N_j\left(\pm t_s-E_{4M}^{(j)}\right)}, \\
	T_B&=\pm\frac{t_w^3}{2}\left(\sum_{j=0}^3\frac{a_3^{(j)}}{N_j\left(\pm t_s-E_{4M}^{(j)}\right)}\right)^2.
	\end{aligned}
	\end{equation}

	\clearpage

\end{document}